%% file: sigma-arxiv.tex
\PassOptionsToPackage{usenames,dvipsnames}{xcolor}
\documentclass[sigconf, nonacm]{acmart}
\usepackage{balance}  
\usepackage{mathtools}
\usepackage[shortcuts]{extdash}
\usepackage[utf8]{inputenc}
\usepackage{csquotes}
\usepackage{IEEEtrantools}
\usepackage{textcomp}
\usepackage{lstautogobble}
\usepackage{suffix}
\usepackage[moderate]{savetrees}
\usepackage{hyperref}
\usepackage{zi4}
\usepackage[font=normal,skip=4pt]{caption}
\usepackage{xspace}
\usepackage{graphicx} 
\usepackage{xcolor}
\usepackage{enumitem}
\graphicspath{{images/}} 
\AtBeginDocument{%
  \providecommand\BibTeX{{%
    \normalfont B\kern-0.5em{\scshape i\kern-0.25em b}\kern-0.8em\TeX}}}

\newcommand\vldbdoi{XX.XX/XXX.XX}

\newcommand\vldbavailabilityurl{} 

\newif\ifarxiv
\arxivtrue

\newcommand{\attr}[0]{$*$}
\newcommand{\screenshot}[1]{\fbox{\includegraphics[width=0.97\linewidth]{#1}}}
\WithSuffix\newcommand\screenshot*[1]{\fbox{\includegraphics[width=0.95\textwidth]{#1}}}

\definecolor{bluekeywords}{rgb}{0.13, 0.13, 1}
\definecolor{greencomments}{rgb}{0, 0.5, 0}
\definecolor{redstrings}{rgb}{0.9, 0, 0}
\definecolor{graynumbers}{rgb}{0.5, 0.5, 0.5}
\definecolor{darkgray}{gray}{0.35}
\newcommand{\level}[1]{\text{\scriptsize\textcolor{darkgray}{#1}}}

\lstdefinelanguage[]{SigmaSQL}[]{SQL}{
  morekeywords={with,over,datediff,nest,unnest,rank,window,lead,lag,nth_value},
  deletekeywords={date},
}
\lstdefinestyle{sql}{
  basicstyle=\ttfamily,
  language=SigmaSQL,
  mathescape=true,
}

\lstdefinelanguage{Sling}{
  classoffset=0,
  keywords={True, False, Null},
  classoffset=1,
  keywords={
    BinRange,
    Count,
    CountDistinct,
    CountIf,
    CumulativeSum,
    DateDiff,
    DateTrunc,
    FillDown
    If,
    IsNull,
    Lag,
    Max,
    MovingAvg,
    Null,
    PercentileCont,
    Round,
    Sum,
  },
  sensitive=false,
  string=[b]",
  alsoletter={[]},              
}

\newcommand{\sql}[1]{\texttt{#1}}
\newcommand{\excel}[1]{\textsf{#1}}
\newcommand{\sling}[1]{\texttt{#1}}
\newcommand{\walg}{\textsf{walg}\xspace}
\WithSuffix\newcommand\subparagraph*[1]{\par \noindent \sloppy \textbf{#1.}}

\newcommand{\ignore}[1]{}
\newcommand{\subhead}[1]{\vspace{0.3\baselineskip}\noindent\textbf{#1}}

\newcommand{\code}[1]{\texttt{#1}}
\definecolor{tomato}{rgb}{1,0.2,0}
\definecolor{turqoise}{rgb}{0.03, 0.91, 0.87}
\definecolor{grey}{rgb}{0.4,0.4,0.4}
\newif\ifnotes
\notesfalse
\newcommand{\cagatay}[1]{\ifnotes{\small[\textcolor{grey}{\c{C}a\u{g}atay:}\textcolor{tomato}{#1}]}\fi}
\newcommand{\jlg}[1]{\ifnotes{\small[\textcolor{grey}{jlg:}\textcolor{tomato}{#1}]}\fi}

\setcopyright{none}
\makeatletter
\def\@copyrightspace{\relax}
\makeatother

\begin{document}

\title{Sigma Worksheet: Interactive Construction of OLAP Queries}

\author{James Gale}
\affiliation{%
  \institution{Sigma Computing}
}
\email{jlg@sigmacomputing.com}

\author{Max Seiden}
\affiliation{%
  \institution{Sigma Computing}
}
\email{max@sigmacomputing.com}

\author{Gretchen Atwood}
\affiliation{%
    \institution{Sigma Computing}
}
\email{gretchen@sigmacomputing.com}

\author{Jason Frantz}
\affiliation{%
    \institution{Sigma Computing}
}
\email{jason@sigmacomputing.com}

\author{Rob Woollen}
\affiliation{%
    \institution{Sigma Computing}
}
\email{rwoollen@sigmacomputing.com}

\author{\c{C}a\u{g}atay Demiralp}
\affiliation{%
  \institution{Sigma Computing}
}
\email{cagatay@sigmacomputing.com}

\renewcommand{\shortauthors}{Gale and Seiden, et al.}



\input{abstract}
\settopmatter{printacmref=false, printfolios=true,printccs=false}
\renewcommand\footnotetextcopyrightpermission[1]{}


\begin{teaserfigure}
  \fbox{\includegraphics[width=\textwidth]{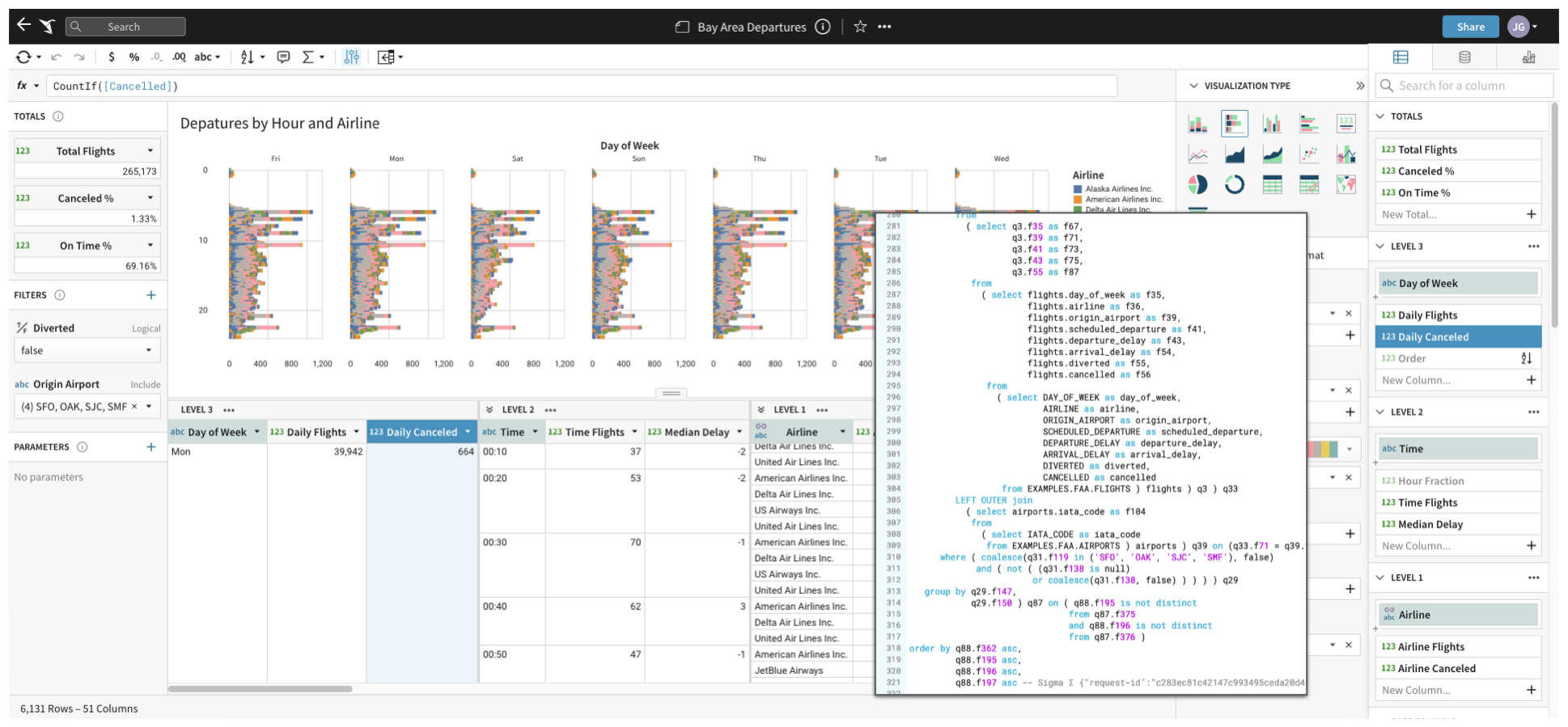}}
  \caption{Sigma Worksheet is an interactive, direct manipulation interface for constructing data warehouse queries---the overlay shows a sample synthesized query. It enables users to access much of the expressiveness of SQL using their knowledge of electronic spreadsheets. Sigma Worksheet is useful for performing ad-hoc analysis, generating complex visualizations, and creating reusable views.\label{fig:teaser}}
  \Description{A screenshot of the Sigma Worksheet interface with an overlay of generated SQL code.}
\end{teaserfigure}

\maketitle



\input{intro}

\input{iterations}

\input{considerations}

\input{overview}

\input{interface}

\input{compiler}

\input{examples}

\input{performance}

\input{feedback}

\input{discussion}

\input{related}

\input{conclusion}

\input{ack}

\balance
\bibliographystyle{ACM-Reference-Format}
\bibliography{sigma}
\end{document}

%% file: abstract.tex
\begin{abstract}
The new generation of cloud data warehouses (CDWs)  brings large amounts of data and compute power closer to users in enterprises. The ability to directly access the warehouse data, interactively analyze and explore it at scale can empower users to improve their decision making cycles.  However, existing tools for analyzing data in CDWs are either limited in ad-hoc transformations or difficult to use for business users, the largest user segment  in enterprises.

Here we introduce Sigma Worksheet, a new interactive system that enables users to easily perform ad-hoc visual analysis of data in CDWs at scale. For this, Sigma Worksheet provides an accessible  spreadsheet-like interface for data analysis through  direct manipulation. Sigma Worksheet dynamically constructs matching SQL queries from user interactions on this familiar  interface, building on the versatility and expressivity of SQL. Sigma Worksheet executes constructed queries directly on CDWs, leveraging the superior characteristics of the new generation CDWs, including scalability.

To evaluate Sigma Worksheet, we first demonstrate its expressivity through \ifarxiv{two real life use cases, cohort analysis and sessionization}\else{a real life use case, cohort analysis}\fi.  We then measure the performance of the Worksheet generated queries with a set of experiments using the TPC-H benchmark. Results show the performance of our compiled SQL queries is comparable to that of the reference queries of the benchmark.  Finally, to assess  the  usefulness of Sigma Worksheet in deployment,  we elicit feedback through a 100-person survey followed by a semi-structured interview study with 70 participants. We find that  Sigma Worksheet is easier to use and learn, improving the productivity of users. Our findings also suggest Sigma Worksheet can further improve user experience by providing guidance to users at various steps of data analysis.
\end{abstract}

%% file: intro.tex
\section{Introduction}\label{sec:intro}
More than twenty years have passed since the introduction of data warehousing~\cite{devlin1988architecture,inmon1992building,chaudhuri1997overview}, multidimensional data cubes~\cite{harinarayan1996implementing,gray1997data},
and business intelligence (BI) tools optimized for these systems~(e.g.,~\cite{aiken1996tioga,stolte2000polaris,stolte2003multiscale}).
In that time, a significant research and development effort  has gone into improving
the online analytical processing (OLAP) ecosystem. Traditional data warehouses  were deployed in data centers on dedicated,
optimized hardware~\cite{inmon2005building}; modern data warehouses are Software-as-a-Service (SaaS)
cloud systems~\cite{Dageville:2016} that require users to do little more than load data, issue queries,
and dial in the balance of speed-vs-cost. Similarly,  traditional BI  systems~\cite{chaudhuri2011overview}  for data exploration and visualization required up-front modeling in
the data warehouse or specialized ``analytics servers,'' in order to interface with tools optimized for querying multidimensional data cubes. Many traditional BI systems such as Power BI~\cite{powerbi}, Tableau~\cite{tableau}, and QlikView~\cite{qlik} have
evolved to have significant improvements in user experience, interactivity, visualization, and dashboarding. While originally designed for on-premise use as dedicated applications, nowadays these systems invariably have cloud offerings, reflecting the trend towards enterprise cloud computing.

Cloud computing has dramatically increased the scale of information within enterprises of every size by reducing the cost of storage while increasing the elasticity. Analysts in these organizations find, however, themselves overwhelmed as they manage an ever growing set of data
sources~\cite{fivetran-survey}. A much larger population of business users (e.g., product managers, operations directors, customer success managers, marketing campaign managers, account managers, etc.), who have limited time or proficiency for writing  queries,  also wish to make use of this data without going through analysts. Although many of  these users are comfortable doing analysis using spreadsheets,  enterprise data often resides in a data warehouse, as it is often the only practical solution for companies to store  billions of
records from their numerous data sources with reliability and compliance guarantees. While many BI systems are well suited to these data sources, spreadsheets are not, suffering from limited scalability and expressivity~\cite{bendre2015dataspread,bendre2018towards,rahman2020benchmarking}. This requires spreadsheet users to invest in new skills to use,
which can be difficult and time consuming to learn and productively use, or rely on BI analysts.

In this paper, we introduce Sigma Worksheet (Worksheet for short), an interactive SaaS system (Figure~\ifarxiv{fig:teaser}\else{\ref{ss:overview}}\fi) that provides a scalable and expressive yet easy to use means
to enable interactive, collaborative visual analysis on datasets that reside in cloud data
warehouses (CDWs). The premise underlying the design of Sigma Worksheet is that supporting the data cube
operations such as slicing, dicing, pivoting, roll-up, and aggregation through an intuitive spreadsheet-like
direct manipulation interface would address the needs of both analysts and business users~\cite{Jagadish:2007}.

Sigma Worksheet is designed as a cloud application and optimized for analyzing data in CDWs It aims to effectively combine accessibility (ease of use and learning), expressivity, and scalability in order to support iterative visual data exploration of  cloud scale data through ad-hoc transformations.  To enhance accessibility, Worksheet integrates an easy-to-use, intuitive
interface with affordances that have made spreadsheet applications successful~\cite{nardi1990}, including a simple
expression language embedded in a table of values, easy
references, easy refactoring and isolation of errors. Worksheet dynamically compiles data operations interactively specified through this interface into SQL queries, building on the versatility and expressivity of SQL. This doesn't just enable expressivity on par with SQL but also amplifies users' ability to generate complex queries that can be otherwise daunting to manually specify. Sigma Worksheet executes compiled queries on a CDW, directly leveraging the desirable properties of CDWs such as scalability, security (e.g., compliance with regulations such as HIPPA and GDPR), elasticity and reliability~\cite{gupta2015amazon,Dageville:2016}. This direct, interactive interface to the CDW differentiates Worksheet from the architecture of current BI systems. These can produce beautiful visuals but then be limited  when users want to get to the ``row level'' data  behind their dashboards or understand what their data means. Unlike spreadsheet applications, Sigma Worksheet enables users to explore billions of records or terabytes of data. In this sense, Sigma Worksheet is an accumulation of the decades-long ideas proposing to combine the accessibility of spreadsheet-like direct manipulation with the characteristics of database systems that enterprises rely on~\cite{raman1999scalable,witkowski2003spreadsheets,witkowski2005query,tyszkiewicz2010spreadsheet,bakke2011spreadsheet,bendre2015dataspread}.



In summary, we contribute Sigma
Worksheet (Figure~\ifarxiv{\ref{fig:teaser}}\else{\ref{ss:overview}}\fi), a new
direct manipulation interface enabling visual interactive OLAP query
construction\footnote{A demo video of Sigma Worksheet is available at:
\url{https://tinyurl.com/y2fssw5a}.}. We describe the design and implementation of Sigma Worksheet enabling an effective integration of accessibility, expressivity, and scalability. We also contribute a discussion on the evolution of our ideas and prototypes that led to the design of Sigma Worksheet, sharing our experiences and
the lessons learned in the process. We demonstrate the expressivity of Worksheet through \ifarxiv{two use cases, cohort analysis and sessionization}\else{an example use case, cohort analysis}\fi, on a dataset with about 200M records and we evaluate the performance of Worksheet generated SQL on the TPC-H benchmark.
Lastly we evaluate Sigma Worksheet through a survey with 100
participants and a semi-structured interview study with 70
participants. We find that Sigma Worksheet improves the efficiency and effectiveness of their analysis
and decision making cycles, irrespective of their previous exposure to SQL or databases.
Our findings also suggest that
users can benefit from machine guidance, particularly during onboarding, informing future iterations of Sigma Worksheet.

%% file: iterations.tex
\begin{figure}[tbp]
  \centering
  \screenshot{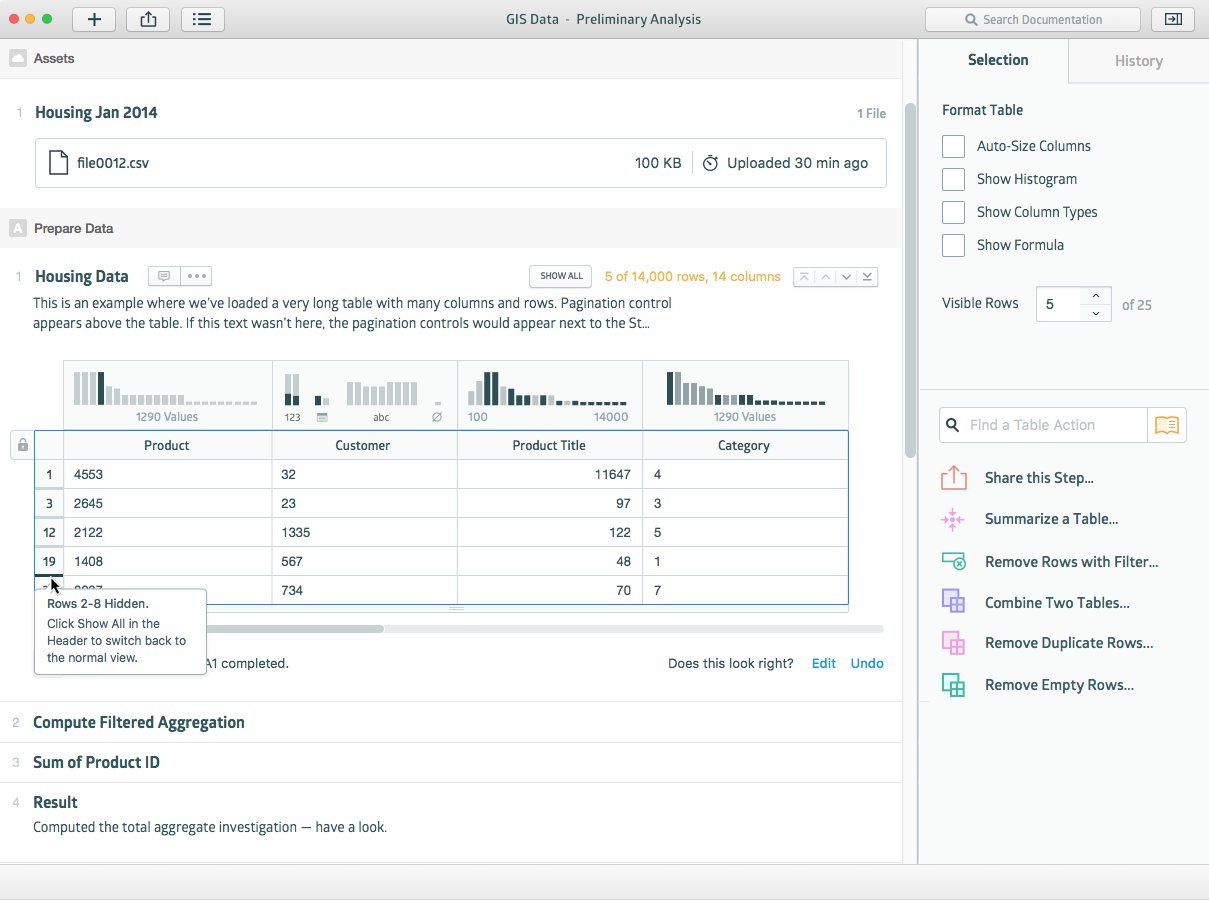}
  \caption{Interface of a notebook-style prototype, one of our earlier
    iterations. This system combined a direct manipulation interface
    with a novel declarative programming language.\label{ss:notebook}
  }
\end{figure}

\section{Iterations and Lessons Learned}\label{sec:iterations}

Sigma Computing is a startup founded with a goal to make data within
organizations more usable to its members. In exploring solutions to
this problem, we built and studied many prototypes which led us to the current Sigma Worksheet. Below we discuss our earlier efforts along with
lessons learned in two categories.

\subhead{Automated Insights} Initially we built several prototype analytic systems whose
functionality might be summarized as ``automatic query generation.’’  These required a semantic
model of the data and so we experimented with different automatic and manual systems for constructing
that model. For data ingestion we built integrations with several SaaS platforms.

As a test case we used sales and operational data from a large technology
company and tried to automatically generate the reports that they were
producing manually. We evaluated these results in interviews with the
individuals who produced the original reports.

From this work we learned that (1)~automatically producing valuable
analyses for our customers was going to require a heavy investment in
up-front modeling, which would make sales difficult and expensive;
(2)~building a high-quality integration for every data source a
customer cared about was daunting for a small company; (3)~capturing
any company specific knowledge with automated methods proved difficult
(e.g., negative sales amounts seemed interesting insight to convey
since they were outliers, but they actually just represented equipment
returns); and (4)~our systems put the business user, along with their
knowledge, on the sidelines. If our users saw something that looked
wrong there wasn't a practical way for them to correct or revise it,
especially if they couldn't understand how values shown to them were
being derived.

\subhead{Visually Programming Data} Reflecting on these prototypes, we
began exploring how to capture and automate the knowledge of business
users. We observed that many contemporary analytics systems such as
Tableau provided an interactive visual interface, but still required
some coding proficiency to be useful in practice. It was our belief
that a sharp drop-off from the visual interface to the programming
layer hampers the productivity of end users. Every time the code needs
a change, the business user becomes blocked on someone else who can
update the code. In the worst case they might have to make do without
that change, leading to incomplete analyses. Interviews with existing
users of these systems confirmed this belief. As such, we asked
ourselves: if we could tightly integrate code with a visual interface,
could we enable business users to become programmers?

Motivated by this question, we implemented a series of prototype
systems that presented data and code together, enabling users to
manipulate both using an interactive visual interface. Our interfaces
were ``live'' in the sense that they automatically refreshed values in response to changes, inspired by Bret Victor's Learnable
Programming\cite{Victor:2012} and electronic spreadsheets. We also
drew ideas from data wrangling systems, including  Wrangler~\cite{wrangler2011} and OpenRefine~\cite{openrefine}, and end user code synthesis approaches such as Excel's Flash Fill~\cite{flashfill}.

We experimented with several interface forms: a spreadsheet-like
canvas, interactive histories of transformations against data, and
notebook-like documents (Figure~\ref{ss:notebook}), similar to Jupyter~\cite{perez2007ipython,jupyter}.
We also experimented with two novel functional programming languages: first
a simple record-based language, and later an array-based language. Both were
influenced by Excel formulas and Microsoft's M language. We investigated the
usability of this platform with our own use-cases and with feedback from
analysts and business users in our target markets.

From the development of this second category of prototypes, we learned (1)
matching a live execution environment with an unconstrained programming
language was difficult, as a small change in the code could lead to an
arbitrarily large change in the data (and in turn the user's view); (2) complex
operations like join and grouping were difficult to integrate into a guided user
interface, alongside arbitrary code; (3) potential customers were reluctant to
entrust direct access to and storage of business data to a cloud-based organization
with no reputation.

%% file: considerations.tex
\section{Design Considerations}\label{sec:considerations}
After our iterations and experimentation summarized above, we set out to build Worksheet, our interactive query builder that would underpin Sigma. We were seeking a middle ground between the two types of systems we had previously explored, a highly-automated system with very limited interaction and a visual programming system.     
We identified five criteria to guide the design and development of
Worksheet. These criteria are informed by what we learned from our
earlier iterations discussed above, existing research and tools for
data analysis (e.g.~\cite{wrangler2011,Victor:2012} and in particular~\cite{Bakke:2016}), and our experience in developing data systems over the years:

\begin{enumerate}
\item[C1] Build on the data warehouse that
  customers already have. Avoid a lengthy ingest phase, endless
  3rd-party integrations and \emph{data controller} 
  responsibility.\label{crit:warehouse:1} 
 
\item[C2] Enable data experts and business users
  to collaborate in a shared language. The interface we offer for
  business users must bring the full power of the underlying database.\label{crit:language:2} 
  
\item[C3] Allow business users to share and
  automate their knowledge. The system must support composition and
  parameterization in a way that is understandable to all of our users. \label{crit:automate:3} 

\item[C4] The query model must be a good match to a live visual interface. The user's view must not change in
  unexpected ways when they are making edits and complex changes must be guided by purpose built interfaces.\label{crit:interface:4}

\item[C5] The query builder must be
  understandable and usable by any user who can use spreadsheet
  systems.\label{crit:usable:5} 
\end{enumerate}

We now discuss the design and implementation of Worksheet. In our
discussion, we refer back to the criteria introduced here as needed to
indicate specific design and implementation choices that they
motivated.

%% file: overview.tex
\begin{figure}[tbp]
  \centering
  \fbox{%
    \begin{minipage}{0.97\linewidth}
    \includegraphics[width=0.97\linewidth]{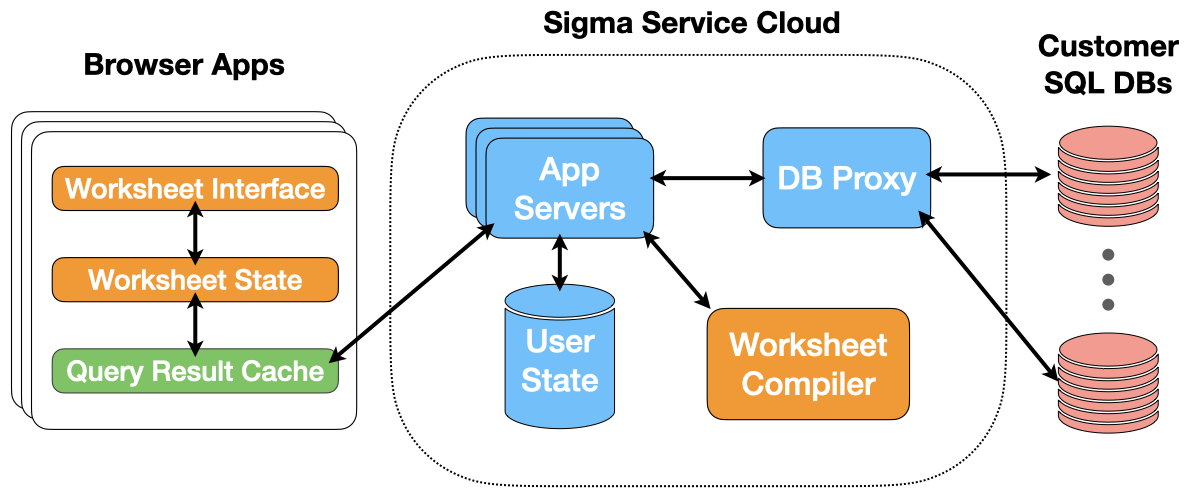} \\[2.0ex]
    \includegraphics[width=0.97\linewidth]{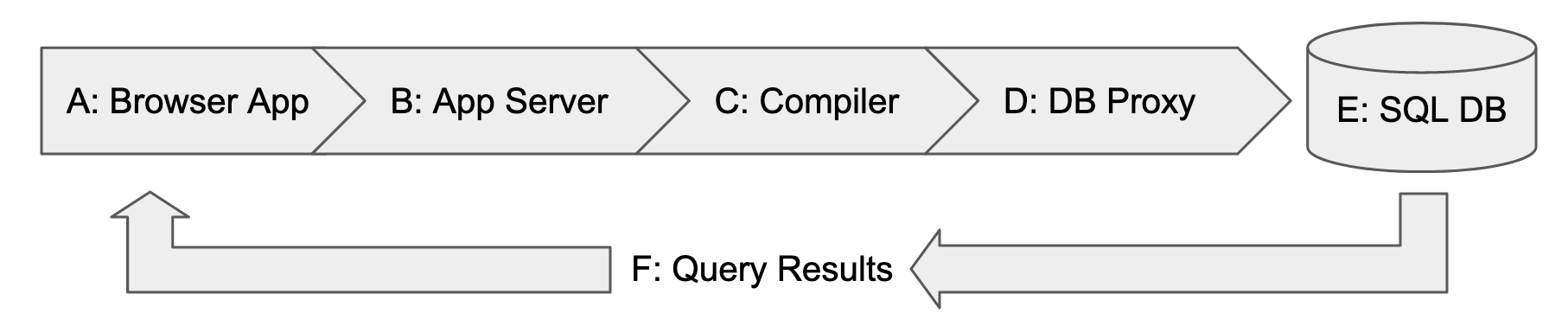}
  \end{minipage}}
  \caption{Sigma architecture (top) and query pipeline (bottom, gray). Sigma is a cloud-based multi-tenant SaaS system for
    interactive visual data analytics. The primary Worksheet elements
    (yellow) are an interface in the browser app (A) and a compiler
    (C) in the Sigma service cloud. Other elements shown support the
    query pipeline which feeds results to the Worksheet interface.
    Sigma's customers configure the service with access to a cloud
    data warehouse that they control (E). The app server (B)
    then acts as an intermediary for query requests that are compiled
    to SQL and a proxy (D) performs workload scheduling and
    interfacing with the database. Database query results (F) are
    delivered directly back to the result cache in the requesting
    browser app and presented to users. \label{fig:architecture}}
\end{figure}

\begin{figure*}[tbh]
  \screenshot*{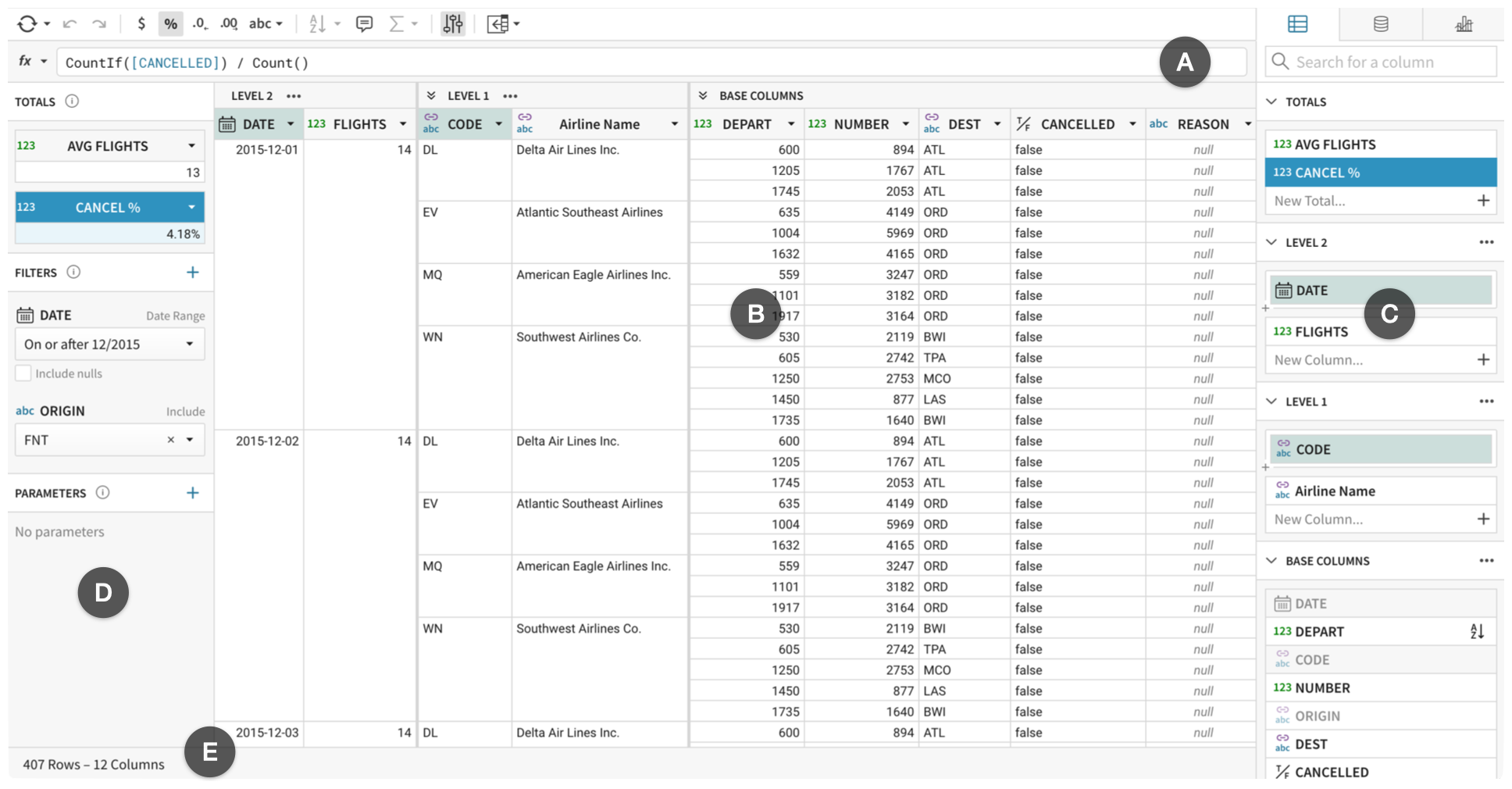}
  \caption{The Worksheet interface. The central feature of Worksheet
    is the data table (B) which displays the results of the query
    described by the worksheet. The level inspector (C) summarizes the
    multi-dimensional hierarchy of aggregation in Worksheet. The
    formula bar (A) enables the inspection of calculated value
    definitions. The control panel (D) displays the filter and
    parameter state of Worksheet as well as ``totals'' values. The result summary (E) gives the
    scale of the data table. Each of these elements supports editing
    through direct manipulation.\label{ss:overview}}
\end{figure*}

\section{System Overview}\label{sec:overview}

Worksheet is a component in Sigma, a multi-tenant
Software-as-a-Service application (Figure~\ref{fig:architecture}). The
primary elements of the implementation of Worksheet are an interface
embedded in the Sigma browser application (Section~\ref{sec:interface})
and a compiler service (Section~\ref{sec:compiler}) which together
support the interactive construction of SQL queries and visualizations
through a direct manipulation, spreadsheet-like interface. The other
architecture elements depicted assist by connecting the Worksheet elements
to each other and with the user's data warehouse in the pipeline shown.


\subhead{Warehouse Integration} Sigma customers configure the service
with access to an RDBMS which they control. No pre-processing or
ingestion is required before users can begin OLAP through Sigma. Sigma
supports multiple warehouse configurations per customer, but as
queries are pushed down to the database, Worksheet is only allowed to
reference data in one warehouse instance at a time.

The elastic and scalable nature of cloud-based data warehouses (CDWs)
enables them to support diverse, interactive OLAP workloads from
large numbers of users. Sigma benefits from these desirable
characteristics by pushing computation to CDWs. It currently supports
BigQuery~\cite{bigquery,melnik2010dremel},
PostgreSQL~\cite{postgresql}, Redshift~\cite{redshift} and
Snowflake~\cite{snowflake,Dageville:2016}.
This interfacing with customer CDWs meets our warehouse
criterion (C1).

\subhead{Query Processing} Access to the customer's data warehouse by the Sigma web application is always mediated by the Sigma service.
Interactive data operations expressed by a user are issued to the
Sigma service as a JSON-encoding of the Worksheet specification
(Section~\ref{sec:compiler}) by HTTP request. The Sigma service
performs authentication, access control checks, query input resolution
and materialized view substitution. The validated, fully resolved
query is sent to the compiler, which generates a corresponding SQL
query.

The SQL query is placed into a workload management queue and
subsequently executed in the customer's database. The results are then
fetched from the database and forwarded directly back to the web
application as-is. Query results with customer data are never stored
in the Sigma service cloud.

\subhead{Query Result Cache} The browser application is structured in
three tiers: (1) interface presentation; (2) application state; and
(3) a query result cache. As changes in the application state
necessitate new query results that don't exist in the cache tier, the
App will dispatch a request to the Sigma service. Query result sets
from the database are returned to the application which updates the
cache. The Query Result Cache matching disregards certain immaterial
differences, such as column name or order, to increase its
effectiveness. By maintaining cache state separate from the
application state, multiple outstanding query requests can be made
without conflict.

\subhead{WYSIWYG} The design of Sigma and the Worksheet is influenced
by ``what you see is what you get'' systems, which enable users to
directly manipulate a view of the final product, rather than
repeatedly modify and recompile code. When Sigma users are making a
complex edit (for instance, joining) that affects the resulting data
in their view, they are provided with appropriate ``previews'' of the
change, to orient them. When users commit a change the Sigma App
refreshes their view to match the specification---issuing database
queries as needed. As such the Worksheet UI never blocks user
interactions while waiting for queries to run. Instead, if an action
would invalidate an in-flight query, the system issues the new query
and signals to the database that it should cancel the prior
one.

\subhead{Interactivity Optimizations} The Sigma system has numerous
optimizations to reduce the interaction latency in Worksheet:
(1)~Sigma caches the database catalog, including table schemas and
constraints; (2)~Sigma prioritizes interactive queries over
offline processing in its workload management queuing. The number of
simultaneous non-interactive queries is limited
such that they cannot overwhelm the warehouse's capacity to accept interactive requests;
(3)~Sigma provides a form of caching so that the results of recent
queries can be re-requested from the customer database;
(4)~the web application maintains a local cache of recent
query results which allows it to avoid re-issuing queries.
(5)~edit manipulations are non-blocking and unneeded queries are cancelled, as described above in WYSIWYG;
(6)~persisted Worksheet specifications may be
configured by the user to be materialized as a table in their CDW on a time-based schedule.
Other Worksheet specifications which reference them will transparently use the table instead of recalculating the result set.

\subhead{Building on the Worksheet} Worksheet specifications are
named, persisted, and reference-able within the Sigma service, similar
to SQL views. Sigma supports collaboration between users through
sharing of these persisted queries both for consumption and as inputs
to other Worksheet specifications, meeting our criterion for
composition and reuse (C3).

Sigma implements a permission system that controls which tables
and documents a user may access. Worksheets can be shared
in a ``viewer'' mode with users who do not have access to the underlying
data sources to authorize them to work with a subset of that data exposed
by the Worksheet.

Result sets from queries constructed using Sigma Worksheet enable
numerous Sigma features, including: (1)~complex visualizations and
pivot tables; (2)~interactive dashboards; (3)~scheduled and
conditional notifications; (4)~embedded visualizations;
(5)~materialized views; and (6)~exporting to spreadsheets.

\subhead{Visualizations}
Supplying and preparing data for visualization is an important application of  
Worksheet. All visualizations in Sigma use a Worksheet specification
for their input. Visualizations in Sigma are displayed in the browser
using Vega~\cite{satyanarayan2016vega} and update in response
to edits of their input Worksheet. Data for the visualizations are
provisioned in the browser app by queries to our app server. The
visualization specification is first translated into a Worksheet
specification and then compiled
(Section~\ref{sec:compiler}). This underscores the expressive power of
Sigma Worksheet, as the expressivity of our visualization interface is
on-par with that of existing commercial visualization products.

\subhead{Worksheet} Sigma Worksheet addresses the challenge of making
database querying easier for non-experts while keeping the expressive
of power of SQL intact for experts~\cite{Jagadish:2007}. For this,
Worksheet (Figure~\ref{ss:overview}) presents an interface similar to
a spreadsheet. Formulas are composed one at a time and can reference
any column in a worksheet, regardless of its position in the
dimensional hierarchy. Input columns can be referenced from anywhere
as well. When a column is renamed, references to it are updated
automatically. Columns may be hidden by the user. Hidden columns do
not display in the data table but may still be referenced within the
worksheet. Hidden columns cannot be referenced from outside of the
worksheet.

The dimensional hierarchy of a worksheet can be directly manipulated
by the user with a pointing device; dragging keys and aggregates to
the desired level or to create new levels. Many elements of the
interface carry context menus that automate common manipulations.

When OLAP is the goal, the product of a worksheet is typically a
collection of visualizations; however, worksheets can support more
than just OLAP. A worksheet is analogous to a SQL View, in that it
exposes a relational schema that is dictated by an underlying query.
Like SQL, the worksheet always has at least one input: this may be a
database table or view, an inline SQL query, or another worksheet. The
last case is particularly important, as it shows that worksheets are
composable building blocks. This is instrumental in attaining our
goals of collaboration, expressivity, and reusability (C2 and 
C3), since it makes worksheets useful for more than
just OLAP: A worksheet may wrangle a JSON document into a table with a
well-defined schema; it may construct a cube-like model to improve
downstream OLAP usability; it may apply row-level permissions to
control which tuples are available to other users.

In the following two sections, we discuss details of the Worksheet
user interface and compiler.

%% file: interface.tex
\begin{figure*}[tbp]
  \screenshot*{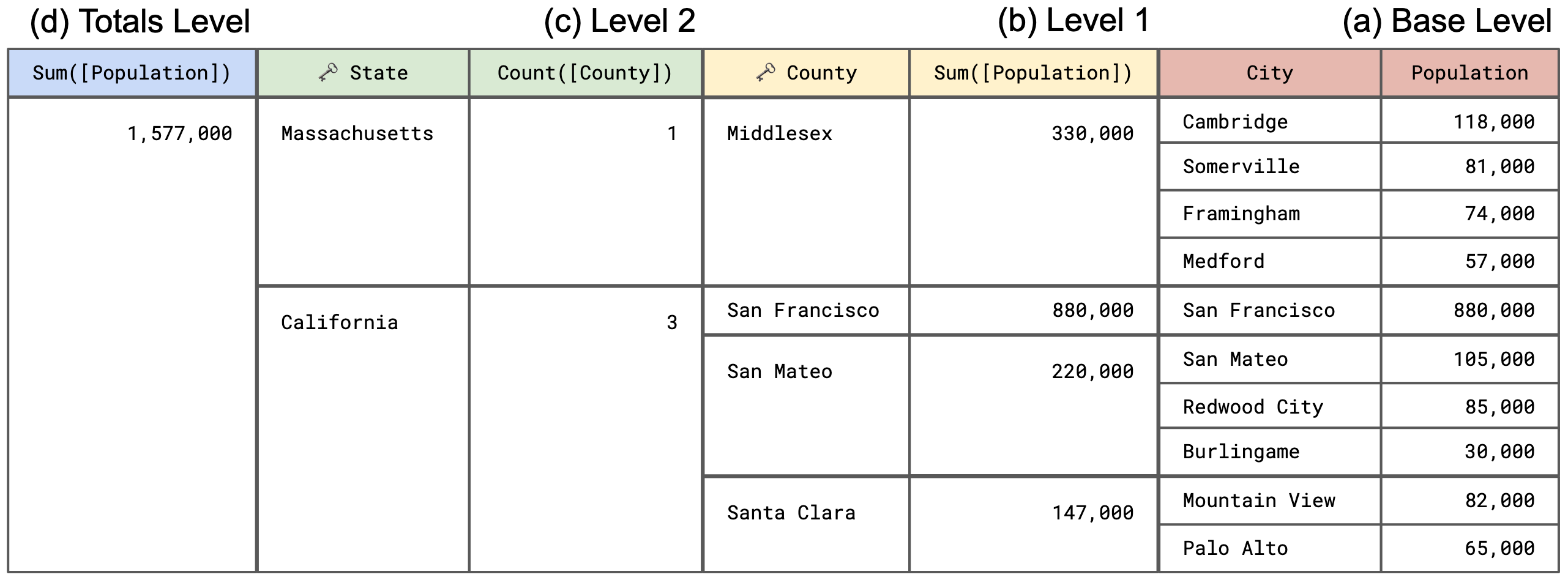}
  \caption{Nested table with four levels, built on a data source
    with four columns: State, County, City, and Population. (a) The base
    level contains City and Population, and is ordered by Population
    in descending order; (b) Level 1's key is County and contains a
    column that aggregates Population; (c) Level 2's key is State and
    contains a column that aggregates County. Note that the count is 3
    and not 6, as County is resident in Level 1 and not the Base
    Level; (d) the Totals Level has a column that aggregates
    Population.\label{ss:levels_diagram}}
\end{figure*}


\section{Worksheet Interface}\label{sec:interface}
The Sigma Worksheet interface is designed with accessibility in mind,
incorporating successful features of spreadsheets in order to enable
business users as well as analysts to interactively explore and
analyze tables in relational databases.
Worksheet translates users' data interactions through direct manipulation into a specification and compiles this specification into SQL (more on this
in Section~\ref{sec:compiler}). As such, our interface design choices are
influenced by spreadsheets and business intelligence products, as
these systems are widely adopted and used successfully by users with
little or no SQL experience.

The Worksheet specification has three primary components: (1)~the
grouping levels within the table; (2)~the formula definitions for
columns in the table; and (3)~zero or more filters applied to the
table. The specification also describes the input data sources and
various data formatting options. These components are sufficient to
meet our shared language criterion (C2). 

We describe them in detail below. \jlg{The data table is conspicuously absent from
  this list. It seems like an important part of the interface. Should
  we discuss pagination? It ties into the optimization discussion
  later.}

\subsection{Levels}
In order to perform aggregation and window calculations, users need a
mechanism to specify grouping and sorting in the Worksheet. In SQL one
would use the \sql{GROUP~BY}, \sql{ORDER~BY}, and \sql{OVER} clauses
to specify this behavior, whereas in a spreadsheet the formula
\code{SUM(A2:A15)} is sufficient. In the Worksheet, users define a
list of grouping levels (level for short) that visually arrange
records in a nested table. A level specifies a set of grouping keys,
map of columns, and an ordering annotation---this enables
expressions such as \sling{CountDistinct} or \sling{MovingAverage} to
derive grouping and ordering properties during compilation.
Furthermore by coupling the level specification to the visual layout
of the table, users can view the inputs and outputs of calculations
side-by-side. For example a user can validate that
\sling{Sum([sales])~$\coloneqq$ 15} by inspecting the input values
\sling{[sales]}~$\coloneqq \sling{[1,2,3,4,5]}$ in a lower level.

Levels in the Worksheet are organized in a hierarchy, as shown in
Figure~\ref{ss:levels_diagram}. The Worksheet always has at least two
levels. The lowest level is known as the \textit{base}, and initially
contains only columns that reference the input data source. It is also
the only level that does not have keys, and therefore is not
aggregated. The highest level, known as the \textit{totals}, has an
empty key set and is used to calculate scalar aggregates. There can be
zero or more levels between these two---in practice we've seen users
construct Worksheets with more than 10 levels. The only restriction on
these levels is that their keys must reference columns from a lower
level.

Levels are similar to a multidimensional \sql{ROLLUP}. This is
because the ``true'' keys of a given level are a union of its own keys
with the keys from all higher levels. For example, if a Worksheet has
three levels (excluding the base) with keys
$\left[\{k1\},\{k2\},\{k3\}\right]$ the ``true'' keys for each level
are $\left[\{k1,k2,k3\},\{k1,k2\},\{k1\}\right]$. This ``cumulative
keyset'' property is important for query generation and optimization
(Section~\ref{sec:compiler}).

The keys of a level are not restricted to expressions that reference
the Worksheet's data source. An expression such as
\sling{Count(\hspace{0pt}[userid])} defined in level 1 could be specified as a key
of level 3. These ``computed level keys'' enable users to visually
express complex sub-queries containing aggregates or windows, using
spreadsheet-like formulas.

Levels can be ``collapsed'' by the user to hide their values---along
with those of lower levels---when no longer needed. Enabling Worksheet
users to view ``grouped'' and ``ungrouped'' values side-by-side
(Figure~\ref{ss:levels_diagram}) facilitates their understanding of the
aggregation and supports our interface criterion (C4).

\subsection{Columns}
A column in Worksheet is defined by an expression, its visibility,
and its ``resident level.'' For example, in Figure~\ref{ss:levels_diagram} the expression \sling{Count([County])} is
resident in level 2. Column expressions, known as formulas, are
written in an expression language familiar to users of spreadsheet and
business intelligence tools. Our supported functions fall into one of
three categories---single row, aggregate, and window---and have
the same behavior as their SQL counterparts. However much like a
spreadsheet there are no restrictions on how these functions compose:
a (convoluted) expression such as \sling{Sum(x + Min(y + Max(MovingAverage(z))))}
is allowed by the Worksheet.

The residency property is used to derive the grouping and sorting for
aggregate and window functions. It is also used when expressions in
one level reference expressions in another level. The most basic
example is an aggregation: a column \sling{[Sales~-~Sum]~$\coloneqq$
  Sum([Sales])} is resident in level 1 and references \sling{[Sales]}
from the base level. Expressions can also reference columns from
higher levels. An example of this is a percent of total: a column
\sling{[\%~Total]~$\coloneqq$ [Sales] / [Sales~-~Sum]} is resident in
the base level and references \sling{[Sales~-~Sum]} from level 1. An
expression is not restricted in how many distinct levels (higher or
lower) it references.

Expressions that reference across levels typically require JOINs in
our generated SQL---we can use the ``cumulative keyset'' property to
derive these JOINs, as this ensures that we have join keys for each
pair of levels. For instance, the percent of total above can join
level 1 and the base using level 1's key. This property also makes
functional dependencies obvious to our compiler, as the key for level
functionally determines the values of \sling{[Sales~-~Sum]}. The
flexibilty to write arbitrary formulas with intuitive references is an important contribution to our usability criterion (C5).

\subhead{Automatic Aggregation}\label{sh:autoagg}
It is common for users to place columns without aggregate expressions
into levels, which have grouping keys. Most databases will return an error
when given a SQL query with the same property. Sigma Worksheet
compensates by collecting the expression's non-null values into a set
and applying the following rules: (a) if the set is empty return
\sql{NULL}; (b) if the set contains a single element return it; (c)
if the set contains 2 or more elements warn the user that there are
multiple values. Rules (a) and (b) ensure that functional dependencies
are preserved on a per-record level, when users omit an aggregation,
while rule (c) provides users with feedback that they've selected the
wrong level keys or column expression. In practice these rules are
implemented efficiently by comparing the minimum and maximum values of
the record group, and does not require materializing the full set.

\subsection{Filters}
The Worksheet provides specialized filter widgets that apply a
predicate to a column's values to select specific records from the
data table. These widgets include: a list of values to
include/exclude; a range of values to include; a simple SQL
\sql{LIKE} pattern to match and a ``top-n'' ranking to apply. For more
advanced predicates users can manually craft Boolean expressions. Some
of these widgets also provide basic statistics and histograms to
assist the user as they configure these widgets. The primary challenge
presented by filters is determining how to order the application of
predicates relative to the calculation of aggregates and windows.
Since the Worksheet does not allow users to specify an explicit order
of operations, we must use a behavior that is explainable and
predictable to our users. Worksheet we employs a ``greedy'' approach to
filter application: as soon as a filter expression's dependencies are
met the predicate is applied.

\subsection{Data Sources}\label{ss:data-sources}
Every Worksheet specification has a primary data source, being either
a database table, a SQL query, an uploaded CSV file or another
Worksheet. Additional inputs can be included from the same types of
sources via joins. Sigma Worksheet can also model fact-dimension relationships
between data sources, using a feature called Links. This feature
enables users to incorporate related data without specifying joins,
and instead rely on the system to generate the appropriate SQL using
the premodeled metadata.

A Worksheet specification may be parameterized with named scalar
values, which can in turn be referred like other columns in formulas.
The default parameter values may be overridden when the Worksheet
is referenced.
All together these capabilities are motivated by our composition criterion (C3).


%% file: compiler.tex
\begin{figure*}[tbp]
  \includegraphics[width=\textwidth]{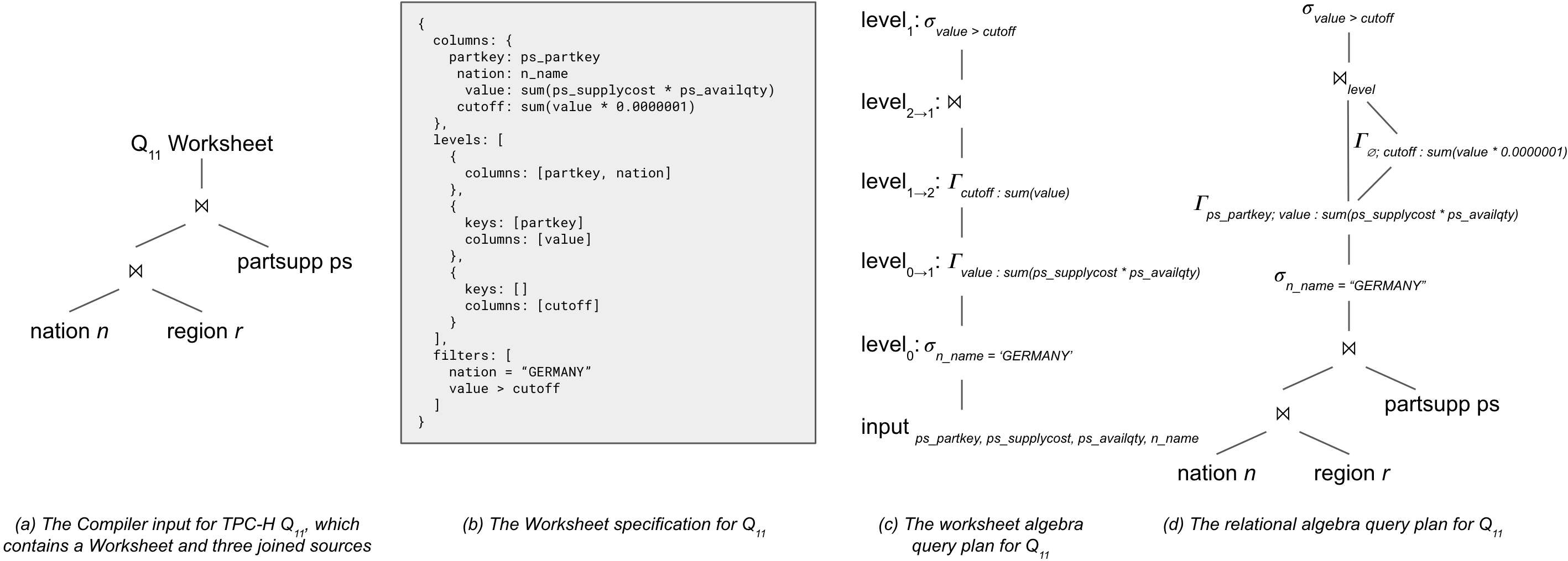}
  \caption{Steps for compiling the worksheet for TPC-H Q11 into a
    relational operator tree. The example query consists of a
    worksheet specification and a data source with three tables (a). The
    query is encoded a schema corresponding to the elements of the Worksheet
    interface (b). The specification is transformed into a Worksheet Algebra
    specification describing the order of operations in the worksheet (c).
    The Worksheet Algebra specification is lowered into a relational operator
    tree that can compile to SQL (d).
\label{fig:compiler}}
\end{figure*}

\section{Compiler}\label{sec:compiler}
Users of Worksheet rely on Sigma to generate correct, optimized
queries in their database's SQL dialect. At a high level our compiler
translates a relational operator tree into SQL, which poses
many practical challenges as the databases we support all
deviate from and/or extend conventional SQL. The Worksheet
specification presents us with further practical challenges, as there
are many non-trivial mappings from the levels, columns, and filters to
a relational operator tree. This has led us to break compilation up
into phases that address specific problems: lowering the Worksheet's
calculation graph into our novel Worksheet Algebra; lowering the
Worksheet Algebra plan into standard relational operators; generating
readable queries in the target database's SQL dialect
(Figure~\ref{fig:compiler}). We also apply rewrites and optimizations
in each compilation phase in order to improve the efficiency of the
generated queries.

\subhead{Calculation Graph} The levels, columns, and filters of a
Worksheet specification form a spreadsheet-like calculation graph.
The column expressions form nodes while function arguments form the
edges. Additional edges are added between expressions in levels and
the levels' keys. Finally filter predicate expressions are added as
special nodes. One of the preconditions for generating a SQL query is
calculating an ``order of operations'' for a given Worksheet specification. This means
ordering the nodes of the calculation graph. In order to produce a
ordering of the calculation graph, we must detect and handle cycles.
To that end, we identify any nodes in a cycle and replace
them with a literal value that has a special error data type. This
data type behaves much like SQL's \sql{NULL} or spreadsheet errors,
and floods through calculations---we perform this recurrent node replacement in the calculation graph during typechecking. After detecting and removing
cycles we are guaranteed that the Worksheet calculation graph forms
a directed acyclic graph (DAG), thus enabling us to compute an
ordering of its nodes.

\subhead{Worksheet Algebra} An ordering of nodes is not sufficient, as
we ultimately need to produce a relational operator tree. To resolve
this, we developed an informal Worksheet Algebra (\walg) to act as a
bridge between the UI's specification and a relational operator tree.
A \walg plan operates on a nested relation, where each level of nesting
corresponds to a level in the Worksheet. As such all \walg  operators
are unary and operate on a nested relation---this includes our join
operator as we only allow joins between a pair of levels. We describe
the operators below.
\begin{itemize}[itemsep=1ex,leftmargin=1em]
\item[]\textit{Select(Predicate)} A select operator applies a
  predicate to a nested table. If a predicate rejects all records
  nested under a group, the group is also rejected. In this way, a
  select is applied to all levels of the Worksheet, rather than a
  single level.
\item[]\textit{Project(Level$_i$, Expressions)} A project operator
  adds new attributes to a specific level in the nested table.
  Multiple operators are required to add new attributes to multiple
  levels.
\item[]\textit{Join(SourceLevel$_i$, TargetLevel$_j$, Expressions)} A
  join operator makes attributes in a source level available in
  another target level, while evaluating expressions. The behavior of
  the operator depends on whether the source level is nested under the
  target level, or vice versa. When the source level is nested under
  the target, expressions with aggregate functions are evaluated;
  those without apply the automatic aggregation described in
  \ref{sh:autoagg}. When the target level is nested under the source,
  the values from the source level are repeated within each record of
  the target.
\end{itemize}
Our ordering algorithm applies a modified topological sort to the
calculation graph, and outputs a \walg query plan. In addition to
calculating a correct ordering of the graph, the algorithm aims to
group multiple graph nodes into a single \walg operator whenever
possible, as an early optimization. For example we might group
multiple single-row expressions from the same level into a \walg
project, or multiple joins between levels into a single operator. We
discuss optimization of \walg query plans later in this section.

\subsection{Relational Algebra}
Our compiler is able to lower every \walg query plan into a relational
operator tree containing standard operators. The lowering algorithm
constructs one subquery for each level, and performs left joins
between these subqueries for expressions that reference columns from
different levels. As such, the algorithm is primarily concerned with
translating each \walg operator into one or more equivalent relational
operators. Consider a worksheet with 5 levels: a \walg join between
levels 2 and 5 uses level 5's keys as the relational aggregate's
grouping keys, and level 2's records as the operator's input; a \walg
join between levels 4 and 1 uses leverages the cumulative keyset
property generate a relational equijoin using level 4's keys; a \walg
select applies a predicate to every level through a combinations of
relational selects and semijoins.

\subsection{SQL Generation}
At first glance, producing a SQL query string from a relational
operator tree is largely a mechanical process. The standard relational
operators generally correspond to clauses in a standard \sql{SELECT}
block, and complex plans are rendered with inlined subqueries or
common table expressions (CTEs).

The primary complexities stem from ensuring that a Worksheet produces
the same results on the same data source, regardless of the target
database.\jlg{We are implying that we go through some trouble to hide
  the differences in warehouse implementations. Why? Do we allow users
  to access to differentiating features of their warehouse?} This
leads to many implementation challenges, as every database we support
deviates from and/or extends conventional SQL in one way or another.
For example, uniformly supporting date/time and timezones across
databases has required a surprising amount of engineering effort: some
have novel date/time data types; some are missing functions that are
native in others.

Production usage has also surfaced surprising issues related to SQL
generation. For example, one database we target rejects queries with
certain semijoins, requiring us to rewrite semijoins to equivalent
operators. Another database can run its own compiler out of memory
when query complexity exceeds a certain threshold. Perhaps the most
noteworthy of these issues is the handling of common table expressions
(CTEs). Initially we used these to reduce duplication in our generated
queries, as they provide a mechanism to name and reference shared
subqueries. However we soon discovered that some of our targets use
CTEs for more than just syntactic sugar. In at least two databases we
support, CTEs have a meaningful impact on query optimization. This
makes it non-trivial to determine whether or not it is preferable to
reference a CTE or inline it as a subquery.

\subhead{Supported Query Types\label{subhead:qtypes}} SQL is an
expressive database query language, which can fully express the
semantics of our Worksheet. Since our workloads exclusively consist of
read-only analyses, we generate \sql{SELECT} statements that include
clauses such as \sql{FROM}, \sql{WHERE}, \sql{GROUP~BY},
\sql{ORDER~BY}, \sql{LIMIT}, \sql{OFFSET}, \sql{OVER}, \sql{JOIN},
\sql{WITH}, \sql{EXISTS}, and others. In our evaluation of Worksheet
using the TPC-H benchmark (Section~\ref{sec:performance}), we are able
to express 20 of 22 queries---Q21 and Q22 are not yet supported as
Worksheet cannot directly express arbitrary semi-/anti-joins. This
limitation underscores that while the Worksheet can express many
useful SQL queries, it cannot express all queries. Removing
limitations on supported query types is an area of active exploration
in our work.

\subsection{Rewrites and Optimizations}
Our query compiler performs rewrites and optimizations in all three of
its intermediate representations: calculation graph, Worksheet
Algebra, and relational query plan. The goal of these rewrites and
optimizations is threefold: generating readable SQL to ease user
inspection and debugging, minimizing the work a target database's
optimizer needs to do, and compensating for rewrites and optimizations
that are not uniformly applied by databases we target. We discuss some
of these rewrites below.

\subhead{Calculation Graph} A calculation graph is a convenient
representation for performing rewrites such as constant propagation,
compile-time evaluation, and dead code elimination (DCE). Of these we
find DCE to have the greatest impact, as removing an expression from
the calculation graph may result in pruning one or more joins or
aggregations from the generated SQL. Some of our rewrites also take
hints from user interactions. For example, DCE kicks in as a result of
hiding a column or collapsing a level.

\subhead{Worksheet Algebra} The Worksheet Algebra is derived from the
relational algebra, thus enabling us to merge, separate, and reorder
operators without changing the overall query plan's semantics. In some
cases we are able to combine adjacent join operators that apply on the
same pair of source and target levels. This optimization is possible
since each level's keys functionally determine its column's values. In
a simple example, we can rewrite adjacent join operators for
\sling{MIN(x)} and \sling{MAX(x)} into a single operator that computes
both expressions. This can have a large impact on our generated
queries, as it reduces the overall number of aggregations and joins in
the generated SQL---a good heuristic for comparing the relative cost
of two queries. Reordering operators can also increase the likelihood
of finding a plan where we can perform these merges.

Another potential optimization in the Worksheet Algebra is something
we've named ``semijoin elision.'' Since a select operator applies to
all levels, there are sometimes cases where we need to filter tuples
in one level based on the set of tuples in another. This typically
shows up in our generated SQL as an EXISTS clause. However, in some
cases where a join between two levels immediately follows a select
operator, we are able to replace the left join with an inner join and
skip the semijoin altogether. This is possible when the join's
``source'' level is already filtered, and its ``target'' level would
otherwise require a semijoin. Again this optimization reduces the
number of expensive operators in the relational query plan.

\subhead{Relational Operators} The optimization of relational operator
trees is a well researched topic. One common optimization we do not
perform is join reordering. Since our compiler has no visibility into
the internals of the databases we target, trying to develop a cost
model would be both redundant and ineffective. Instead we focus
 our query rewriting on removing occasional redundancies due to earlier compilation phases, by dropping noop operators,
merging adjacent pairs of project or select operators, or pruning
unneeded operator attributes. While these rewrites don't impact query execution, they reduce noise in the generated SQL.

When a Worksheet Algebra plan is lowered to relational operators, mark the relational left joins to indicate they were generated from \walg joins. Although we've lost the metadata describing the level keys,
knowing that the same key constraints apply enables us to perform
rewrites. The first is a final round of join pruning that applies when
a query does not use any attributes from the right leg of an annotated
left join; this rewrite is currently applied during attribute pruning.
The second is a sort-limit pushdown, which enables us to swap an
order-by limit with an adjacent annotated left join. The motivating
example is a join between the base level and an aggregate level, where
the number of records in the base is multiple orders of magnitude
greater than the aggregate level. When the ordering exclusively
references attributes in the base level, we can apply the limit before
the join, thus potentially decreasing the execution cost of the join.
Furthermore some target databases are able to select efficient
execution plans for ordered limits, which can lead to even greater
performance improvements. \jlg{Can we explain that the limit comes
  from Worksheet's interactive nature.}

%% file: examples.tex
\begin{figure}[tbp]
  \centering
  \fbox{\includegraphics[width=\linewidth]{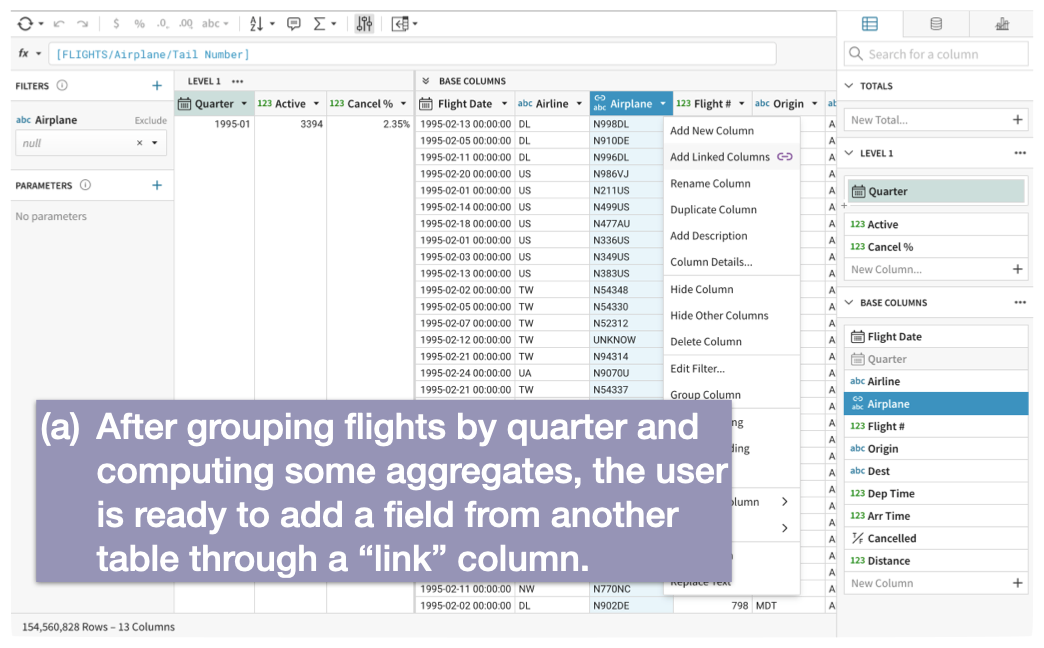}} \\
  \fbox{\includegraphics[width=\linewidth]{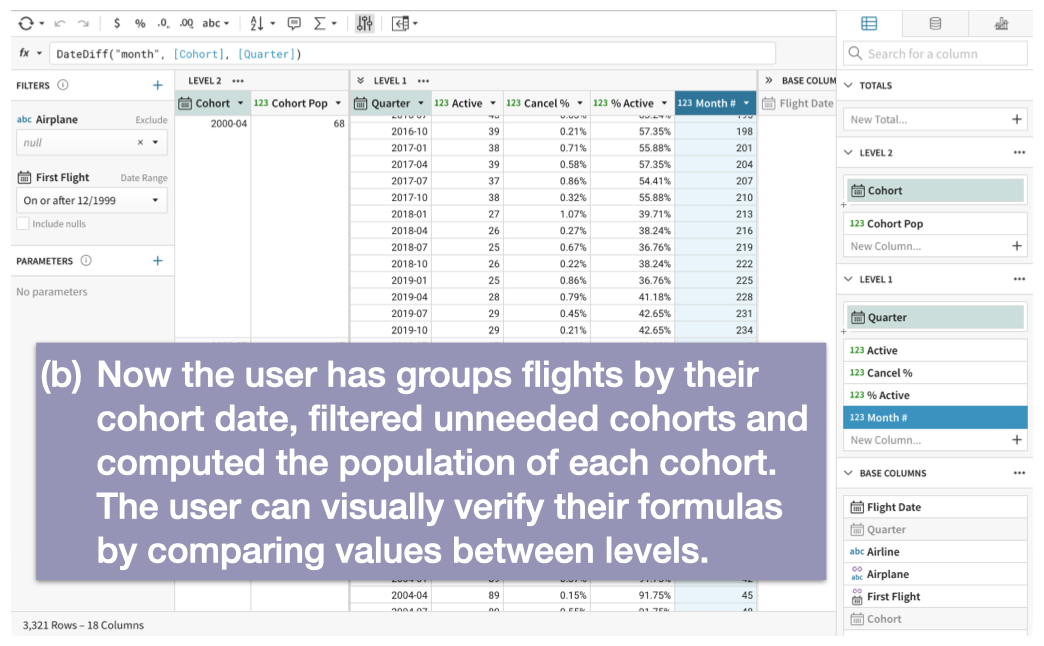}} \\
  \fbox{\includegraphics[width=\linewidth]{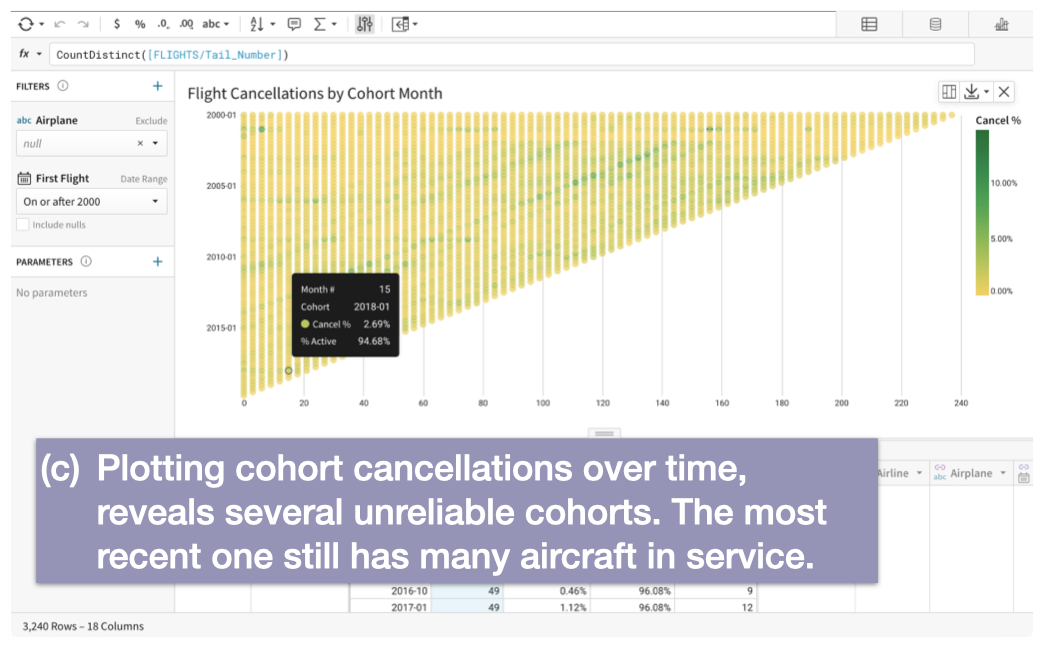}} \\
  \caption{Cohort analysis example: constructing an aircraft reliability
    visualization. The user easily enriched their table by joining in
    another Worksheet specification through a link.\label{example1}}
\end{figure}

\section{\ifarxiv{Examples}\else{Example}\fi}\label{sec:examples}

\ifarxiv{We will now demonstrate the capabilities of Sigma Worksheet
  for common business intelligence analysis with two example
  scenarios}\else{We will now demonstrate the capabilities of Sigma
  Worksheet for common business intelligence analysis with an example
  scenario}\fi. \ifarxiv{Both examples}\else{We}\fi\ utilize the
On-Time database of United States domestic airline carrier flights
between 1987--2020~\cite{transstats:2020}. The main \sql{FLIGHTS}
table is around 200M rows. Each row contains information about a
scheduled carrier flight, including the date, carrier, aircraft
identifier, scheduled times, actual times, and

In \ifarxiv{these scenarios}\else{this scenario}\fi\ an
aircraft-reliability manager is using Sigma Worksheet to perform
analysis. The manager is responsible for making decisions about their
carrier's maintenance policies. They are not familiar with SQL and
before Sigma were accustomed to working with analyst-supplied data
warehouse extracts in Excel.


When necessary we include Sigma Worksheet formula expressions in these
examples. The syntax and semantics are similar to simple SQL
expressions. One difference is that column references are quoted in
brackets (\sling{[]}). The Level where a formula is defined is
indicated by an abbreviation in the subscript, e.g, L1 for Level 1, B
for Base Level, etc.

\subhead{Cohort Analysis} Our manager wants to characterize airplane
reliability among planes which entered service at the same time. This
type of analysis, known as ``cohort analysis,'' is a common
analysis with longitudinal datasets. Sigma Worksheet is useful for
this task, easily creating the analysis in a few simple steps.

In Sigma, database tables and saved Worksheet specifications co-exist
in a browseable hierarchy. Sigma also provides a search engine. The
manager uses the search interface to quickly find the \sql{FLIGHTS}
table which they were previously aware of. Opening the table in Sigma
displays a preview the table's contents along with a prompt to begin
the analysis in Worksheet. Our manager proceeds, performing their
analysis in three steps.

First the manager adds a level grouping flights by quarter year, with the
key shown in Formula~\ref{eq:quarter} and a calculation of active
planes and cancellation percentage, Formulas~\ref{eq:active-planes}
and~\ref{eq:cancel-pct-1} respectively.
\begin{IEEEeqnarray}{rCl}
  \sling{[Quarter]}_{\level{L1}} & \coloneqq & \sling{DateTrunc("quarter",[Flight~Date])}
  \label{eq:quarter} \\
  \sling{[Active]}_{\level{L1}} &\coloneqq& \sling{CountDistinct([Tail~Number])}
  \label{eq:active-planes} \\
  \sling{[Cancel \%]}_{\level{L1}} & \coloneqq & \sling{CountIf([Cancelled])/Count()}
  \label{eq:cancel-pct-1}
\end{IEEEeqnarray}

The manager wants to group planes by ``cohort,'' which in this case
will be the quarter when the plane first flew a scheduled flight. The
first-flight of the plane is not present in the \sql{FLIGHTS} table
but it is accessible through a link (described in
Section~\ref{ss:data-sources}) to a saved Worksheet specification,
titled ``Planes,'' where it is present. The Worksheet indicates the presence
of this link to the manager through a ``link column'' called
\sling{[Plane]} which displays the tail number of the plane. The
\sling{[First~Flight]} column can be quickly added into the table
through a simple drop-down menu (Figure~\ref{example1}a). After adding
the column, the manager adds a filter which excludes flights whose
plane entered service before December 1999, being too old and not
necessary for this analysis.

Now the manager adds a second level above the first, grouping
cohorts with the key in Formula~\ref{eq:cohort} and computing
total planes in the cohort,
Formula~\ref{eq:cohort-planes}. With that they find the
percentage of active planes from the cohort for each quarter in the
first level, Formula~\ref{eq:pct-active}, and the months since the
start of the cohort, Formula~\ref{eq:months} (Figure~\ref{example1}b).
\begin{IEEEeqnarray}{rCl}
  \sling{[Cohort]}_{\level{L2}} & \coloneqq & \sling{DateTrunc("quarter",[First~Flight])}
  \label{eq:cohort} \\
  \sling{[Cohort~Pop]}_{\level{L2}} & \coloneqq & \sling{Max([Active])}
  \label{eq:cohort-planes} \\
  \sling{[\% Active]}_{\level{L1}} & \coloneqq & \sling{[Active]/[Cohort~Pop]}
  \label{eq:pct-active} \\
  \sling{[Month \#]}_{\level{L1}} & \coloneqq & \sling{DateDiff("month",[Cohort],[Quarter])}\IEEEeqnarraynumspace
  \label{eq:months}
\end{IEEEeqnarray}

Finally they create a diagonal heatmap
visualization (Figure~\ref{example1}c) based on this
worksheet data. The visualization
uses the Worksheet columns as its input. The x-axis encodes
the months since the start of
the cohort (Formula~\ref{eq:months}), the y-axis encodes the starting
quarter of the cohort (Formula~\ref{eq:cohort}), and the color encodes
the percentage of cancellations (Formula~\ref{eq:cancel-pct-1}).
The manager observes ongoing, high-cancellation rates for
the cohort that began in 1Q 2018, of which a large portion remains in
service. This discovery leads the manager to open an inquiry with the
aircraft manufacturer.

Our manager was familiar with the ``Planes'' data from the second step
because they had created it themselves previously. It is itself a
Worksheet specification built from the \code{FLIGHTS} table with an
added level keyed by \sling{[Tail~Number]} and a column
finding the first flight, \sling{[First~Flight]}$_{\level{L1}}$~$\coloneqq$ \sling{Min([Flight~Date])}.
They annotated the \code{FLIGHTS} table with the link---in this case
just a mapping between the \sling{[Tail~Number]} in both
datasets---to facilitate
future analysis.

\ifarxiv{
\begin{figure}[tbp]
  \centering
  \fbox{\includegraphics[width=\linewidth]{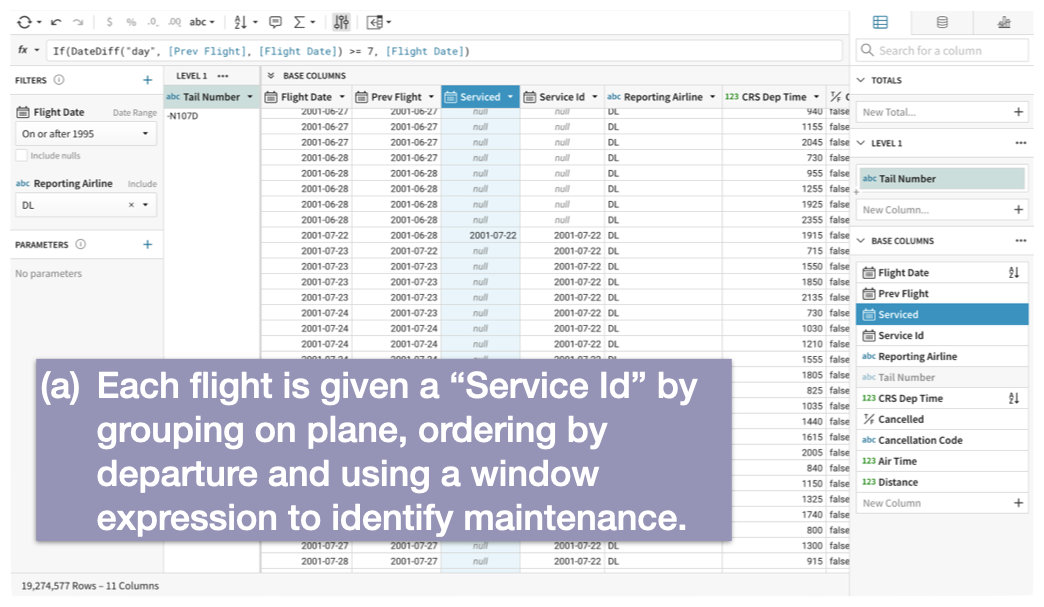}}\\ 
  \fbox{\includegraphics[width=\linewidth]{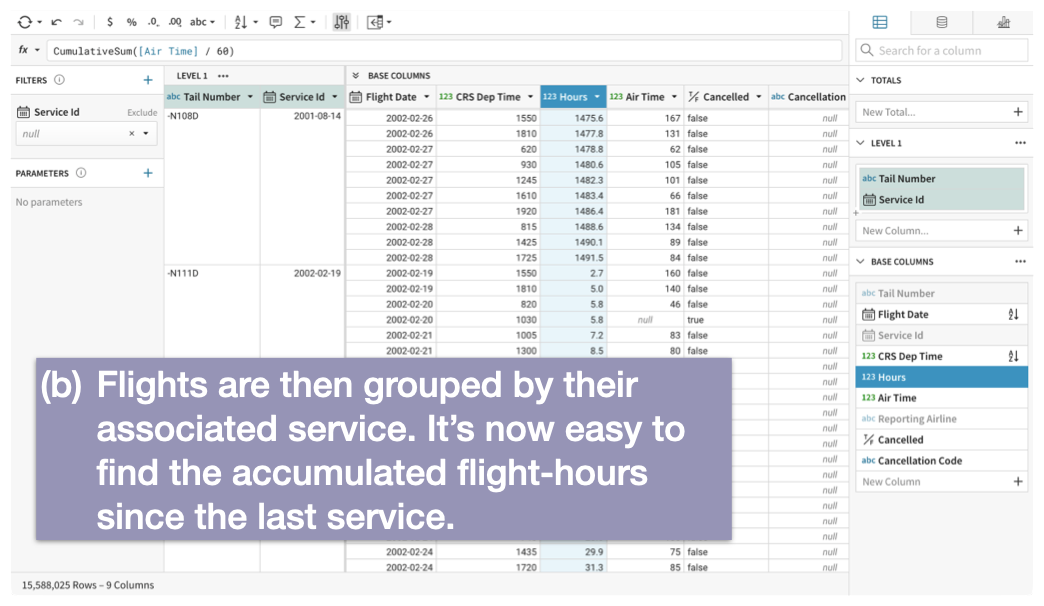}}\\ 
  \fbox{\includegraphics[width=\linewidth]{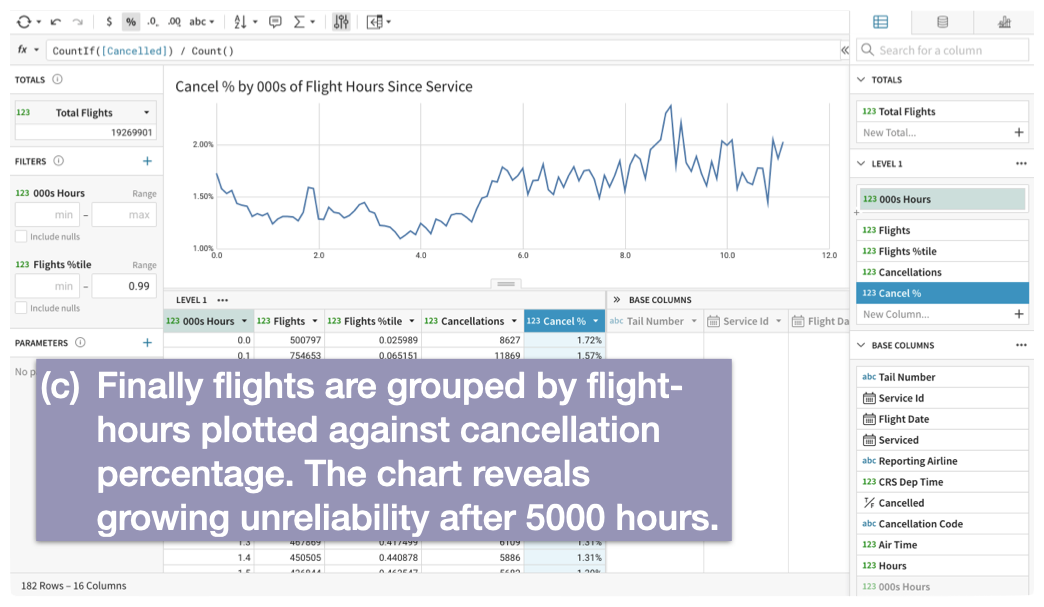}} 
  \caption{Sessionization example: converting flight event data into
    ``sessions'' for a reliability-analysis visualization. Each step
    is grouping the flights records in a different way in order to
    calculate a window or aggregate value. The user can verify the
    window calculations themselves as the \sql{OVER} specification is
    derived from the level specification.
    \label{example2}}
\end{figure}
\subhead{Sessionization}
In another case our manager needs to investigate the effect of
aircraft maintenance checks on aircraft reliability. The On-Time
data does not include maintenance information explicitly, but
major maintenance checks can be inferred from long intervals between
flights. The manager performs an enrichment known as
``sessionization'' to associate aircraft flights with their proceeding
aircraft maintenance checks.

First our manager groups the flights by plane, as in the cohort
analysis. To identify maintenance, they order
the flights of each plane by departure and add two new columns. The
first new column identifies the first flights after a seven day furlough, Formula~\ref{eq:service-completed}; and the second column tags each
subsequent flight with this date as a \emph{session identifier},
Formula~\ref{eq:service-id} (Figure~\ref{example2}a).
\begin{IEEEeqnarray}{rCl}
  \sling{[Prev Flight]}_{\level{B}} & \coloneqq & \sling{Lag([Flight~Date])}
  \label{eq:prev-date} \\
  \sling{[Serviced]}_{\level{B}} & \coloneqq &
  \sling{If(DateDiff("day", [Prev~Flight],} \nonumber\\
    & & \sling{[Flight~Date]) >= 7, [Flight~Date])}\IEEEeqnarraynumspace
  \label{eq:service-completed} \\
  \sling{[Service~Id]}_{\level{B}} & \coloneqq & \sling{FillDown([Serviced])}
  \label{eq:service-id}
\end{IEEEeqnarray}
Referencing values across rows was something the manager did easily
in spreadsheets. Learning to do the same in Sigma Worksheet, using \emph{window functions} like \sling{Lag} and \sling{FillDown},
was less intuitive at first. When the manager is unsure about a function, they
often write the invocation in its own column, as in Formula~\ref{eq:prev-date},
which allows them to ``debug'' when necessary by inspection of the column
values. The behavior of \sling{Lag}, \sling{FillDown} and other window functions
is determined by the
grouping of the levels above it as well the ordering of the level. The
manager needed some help to learn the use of these special functions
as they are different from the more explicit spreadsheet equivalents.

Now for each scheduled flight the manager wishes to compute the
accumulated flight hours since the last maintenance check.
They need to group flights by \sling{[Service~Id]}, but that column
is itself dependant on the current grouping through \sling{Lag}.
The solution is build a new Worksheet
specification with this current one as the input. Then they are free
to re-group columns as desired. The manager does so
(Figure~\ref{example2}b) and after creating a new level with two keys,
\sling{[Tail~Number]} and \sling{[Service~Id]}, the manager again
orders the flights by departure and computes the cumulative flight
hours since the service, Formula~\ref{eq:flight-hours}.
\begin{equation}
  \sling{[Hours]}_{\level{B}} \coloneqq \sling{CumulativeSum([Air~Time]/60)}
  \label{eq:flight-hours}
\end{equation}

The manager now wishes to group flights in a different way, by their
accumulated flight-hours. As before, they know they can't change the
grouping of the flights to \sling{[Hours]} as it depends on the
current level specification. So once again the manger creates a new
worksheet on top of this one, now grouping by a quantization of
\sling{[Hours]} and computes the cancellation percentage
(Formulas~\ref{eq:thousands-hours} and \ref{eq:cancel-pct}).
\begin{IEEEeqnarray}{rCl}
  \sling{[000s Hours]}_{\level{L1}} & \coloneqq & \sling{Round([Hours]/1000,1)}
  \label{eq:thousands-hours} \\
  \sling{[Cancel \%]}_{\level{L1}} & \coloneqq & \sling{CountIf([Cancelled])/Count()}
  \label{eq:cancel-pct}
\end{IEEEeqnarray}

The resulting plot of these columns (Figure \ref{example2}c) shows the
manager a ``bathtub curve'' where some initial unreliability gives way
to an extended period of reliability, followed by growing
cancellations after 5,000 hours of flight. Empowered by this
information, the manager recommends adjustments to the fleet's
maintenance schedule.
}\fi

\subhead{Discussion}
In our \ifarxiv{examples}\else{example}\fi\ the aircraft-reliability manager was working
interactively with a large dataset in their data warehouse. There was
no necessary preparation that could not be done with the combination
of columns, filters and levels offered by Sigma~Worksheet. As they
interacted with Worksheet, the interface continuously provided
feedback with refreshed table and visualization query results. Sigma
Worksheet's simple expression model allowed them to write intuitive
formulas (like Formula~\ref{eq:months}) without being aware of the
complexities of the implementation. These examples demonstrate how Sigma
Worksheet meets our usability criteria (C5).

\ifarxiv{The}\else{This}\fi\ cohort scenario is important enough that investigations have been
made to extend relational database support for it~\cite{jiang2016}.
Sigma Worksheet has no special optimization for cohort analysis, but
it can be expressed simply with a few basic aggregate expressions.
This analysis is also possible in Power BI but requires comparatively
complex DAX formulas~\cite{power-bi-cohort}.

\ifarxiv{
In the sessionization example the user employed window functions
(Formulas~\ref{eq:prev-date}, \ref{eq:service-id} and~\ref{eq:flight-hours}) to
calculate values across grouped rows. They used the levels to define
the ``window'' of related rows. In contrast to SQL where this is done
with an \sql{OVER} clause attached to every function, Sigma Worksheet
window formulas are simpler, at the cost of being less expressive. For
Worksheet users this trade-off is a benefit because the window is
always visible to them in the data table and reference-able as they
compose or debug their formulas. Because of Sigma~Worksheet's single
hierarchy of levels, it is necessary in some cases, as in this
example, to compose multiple Worksheets to complete an analysis.
}\fi

Our \ifarxiv{examples}\else{example}\fi\ required a small amount of data enrichment to complete.
Sigma Worksheet simplifies this task and makes it accessible to the
business user where previously they may have been blocked, waiting for
an analyst or data specialist. Results of this analysis are likely to
raise new questions which may require more understanding of the data.
This makes the self-service design of Sigma Worksheet especially
valuable. The Worksheet user is equipped to immediately go deeper.

%% file: performance.tex
\section{Performance Analysis}\label{sec:performance}

\begin{figure}[tbp]
  \centering
  \includegraphics[width=\linewidth]{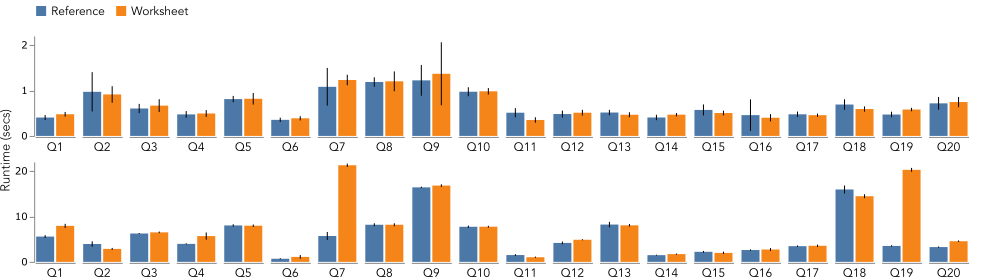}
  \caption{TPC-H query runtime comparison using the SF1 (1GB, top) and SF1000 (1TB, bottom)
    dataset sizes. For each benchmark query, we run the reference SQL
    implementation and the Worksheet generated SQL ten times against a
    Snowflake CDW (2X-Large) and report the mean and the standard
    deviation of the execution times. The performance of the Worksheet
    SQL and the reference SQL is comparable for all but two queries,
    Q7 and Q19. Two of the benchmark queries, Q21 and Q22, are not
    currently expressible in Worksheet. \label{fig:tpch}}
\end{figure}

To assess the suitability of Worksheet to replace hand-written SQL
queries, we conducted a set of experiments using TPC-H-like data and
queries. For twenty of the standard TPC-H queries, we constructed a
semantically equivalent query specification using Worksheet, against
both SF1 (1GB) and SF1000 (1TB) data sets, in one of the
implementations supported by Sigma: a Snowflake \emph{2X-Large}
CDW. We compared the execution runtime of the underlying SQL
generated for each Worksheet specification by our compiler to the
corresponding reference SQL query (Figure \ref{fig:tpch}) to learn if
there is a penalty to constructing queries in Worksheet.

We found that Worksheet SQL is on par with the reference SQL in 18
cases, with the average Worksheet SQL execution in these cases 8.3\%
slower than its counterpart at SF1000 and 0.4\% faster at SF1. Of the
remaining queries, two were not expressible in Worksheet and two were
performance outliers.

Q7 and Q19, showed poor performance in their Worksheet versions, 270\%
and 480\% slower on average, respectively, than the reference SQL at SF1000.
Both cases have the same root cause. Q7 consists of an aggregate over
a join between the \sql{ORDERS} and \sql{LINEITEM} tables (among
others). In the reference case the warehouse query planner was able to
partially push-down~\cite{chaudhuri2000groupby} the aggregate and a
related filter below this costly join. Worksheet's SQL did not benefit
from this optimization. \cagatay{Can we rephrase the following sentence?} Not---we believe---for any semantic reason but
because two extra layers of subquery wrapping in the generated SQL
confused the warehouse query planner. Q19 has a similar aggregate and
join, between \sql{LINEITEM} and \sql{PART}, that failed to optimize
for the same reason. Q21 and Q22 include anti-joins (\sql{NOT~EXISTS~(\ldots)})
which are not yet expressible in Worksheet (see Section~\ref{subhead:qtypes} for details).

The architecture of Sigma incurs some overhead, as query requests must
first be sent to the Sigma service cloud before being forwarded to the
warehouse. To mitigate this Sigma offers its users the choice of
service hosted from Amazon Web Services~(AWS) or Google Cloud
Platform~(GCP) to best align with the user's choice of warehouse
implementation. Compilation of the Worksheet specification into SQL
has a cost as well. On these TPC-H queries we measured an average of
28ms.

%% file: feedback.tex
\begin{figure}[tbp]
  \centering
  \includegraphics[width=\linewidth]{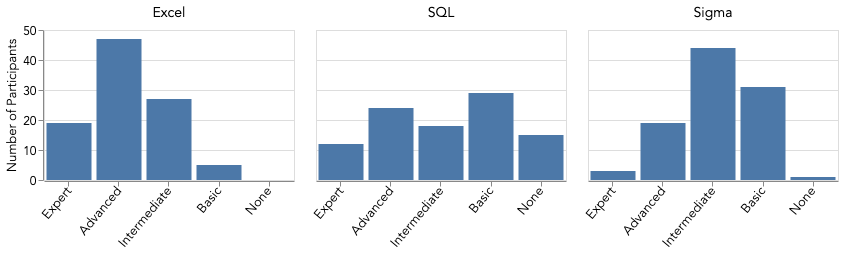}
  \caption{Self reported skills of participants for SQL, Excel, and Sigma. While all participants had some experience with Excel, more than 15\% of participants had no experience with SQL.  The advanced and the intermediate level users were  respectively the largest groups for Excel and Sigma. \label{fig:skills}}
  \end{figure}
  
\section{User Feedback}\label{sec:feedback}
Sigma has an active user base formed by mixed communities of analysts
and business users. Analysts are generally familiar with SQL while business users
are not. Sigma's user base is relatively small but rapidly growing:
over 3,000 monthly active users at the time of writing. We
daily handle more than 250K database queries on their behalf.
Over 25\% of these queries are attributed to ad-hoc analyses
performed by more than 75\% of our user base.

Since Sigma’s first release, we have been regularly collecting
feedback from our users both anecdotally and methodically. Recently we
also conducted a survey study with 100 current Sigma Worksheet users.
We followed up this survey with a semi-structured interview study with
70 participants from the survey in order to elicit detailed feedback
and better contextualize the survey responses. With these two studies,
we aimed to find answers for several questions:
\begin{itemize}[leftmargin=1em,topsep=1ex]
\item[]What high-level goals and expectations do our users have for BI?
\item[]What are their skill levels in Excel, SQL, and Sigma.
\item[]What BI tasks do they perform regularly?
\item[]What are the pain points in these tasks?
\item[]What other BI tools did they use?
\item[]What was their company's reasoning in choosing Sigma?
\item[]How could Sigma better meet their needs?
\end{itemize}
Participants of our study hailed from 35 companies and had job titles
ranging from barista to CEO, with product manager and BI analyst being
the most common. They also had varying degrees of expertise with data
analysis tools (Figure~\ref{fig:skills}). We summarize our findings
below, combining our findings from the survey and the 1-on-1 interviews.

\begin{figure}[tbp]
  \centering
  \includegraphics[width=\linewidth]{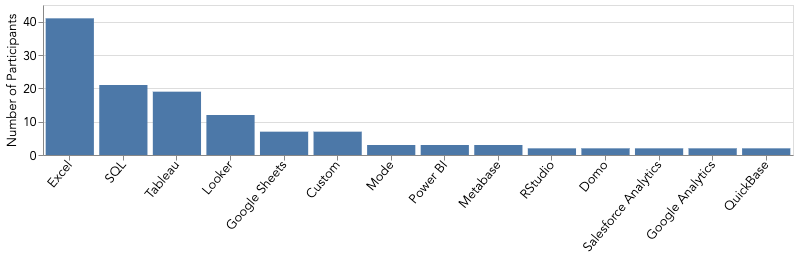}
  \caption{Participant experience with other BI tools. We consider only the tools that were reported by more than one participant.\label{fig:otherbi} \jlg{Can we tighten up the labels somehow?}}
  \end{figure}

\subhead{Prior Experience, Analysis Goals, Pain Points} About half the participants  had some experience with Excel and Google Sheets, one fifth with SQL and Tableau (Figure~\ref{fig:otherbi}). Custom in-house tools were relatively frequent. There was some correlation with prior SQL experience and comfort with using Sigma. Across the board, participants’ overall analysis goals reduced to understanding, tracking, and reporting how their data related to business performance metrics that they cared about.  Example metrics were net sales, new customers, churn rate, retention rate, revenue growth, and return on investment (e.g., of marketing). General analysis patterns included cohort analysis, sessionization, and attribution analysis. Frequently cited pain points with prior BI tools were difficulty of use and collaboration, limited scalability, lacking in ad-hoc transformations, and limited reporting (e.g., visualization) capabilities.

 \begin{figure}[b]
  \centering
  \includegraphics[width=\linewidth]{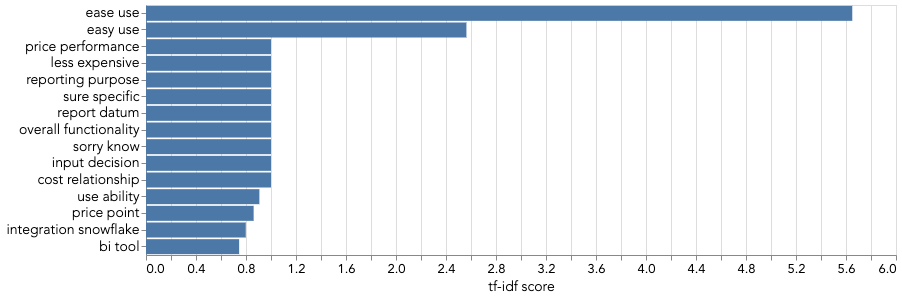}
  \caption{Fifteen most frequent bi-grams (sequences of two adjacent words) found in participants’ responses to why their company chose Sigma.  We compute the  bi-gram frequencies  using the tf-idf weighting~\cite{salton1988term}.
  The survey responses and interviews suggest that customers chose Sigma predominantly for its ease of use. Also integration with Snowflake played a role, as did Sigma’s desirable cost-benefit characteristics.\label{fig:whysigma}}
  \end{figure}

\subhead{Workflow Experience} Participants reported ease of use (Figure~\ref{fig:whysigma}), which helped increase their  productivity, as the most frequent reason for choosing Sigma. They also reported that the live feedback during Worksheet editing was  beneficial, helping them to identify problems earlier. Benefits extended beyond business users. Analysts with SQL experience had understanding of the capabilities of databases. Viewing Worksheet as a type of visual query builder immediately gave them ideas of what they might do with it.

\subhead{Spreadsheet Interface} For business users the database is
very abstract. They often aren't clear about what it can do for them.
We found users' first experience with Sigma Worksheet was through a worksheet
created and shared by a more experienced user. Initially their  usage was learning
how to manipulate columns: extending the existing analysis in a private copy. Eventually they began making their own visualizations. Finally they graduated to creating their own
worksheets to answer their own questions. These users have reported that the spreadsheet-like
interface did make Worksheet more approachable to get started.
On the other hand, some users expected Worksheet to have a complete spreadsheet  semantics, including an editable data grid, rather than a read-only data table. This motivated our ongoing work to enable live data editing in future versions of Worksheet.


\subhead{Data Exploration} We found that the ability to explore data with flexible transformations was one of the main reasons for adopting Sigma. Once they got familiar with Sigma's capabilities, business
users were able to perform extensive exploratory analysis without
relying on BI analysts. To this end, users found the interactive
coordination between Sigma Worksheet and the visualization view (i.e.,
they could have a visualization open and manipulate the data and
observe how the visualization was changing) particularly useful.

\subhead{Scalability} Handling bigger datasets was among the most frequently cited strengths of Sigma. Participants had been dissatisfied in this regard with their
prior direct manipulation tools, spreadsheet applications and BI systems alike. Nevertheless, some participants asked for reducing the query response times with big datasets even more, particularly for further facilitating interactive exploration at scale.

\subhead{User Guidance} While most users found Worksheet easy to learn and use, those primarily with spreadsheet experience were slowed down during onboarding by the differences between the Worksheet and spreadsheet interfaces. Our levels could be also a source of confusion, as some users expected them to have the behavior of a pivot table. Also, new business users occasionally got stuck when executing formulas, creating levels or performing joins. Although most users were able to push through these holdups and became productive, their comments suggested that the Worksheet user experience could benefit from machine guidance and curation at various stages of analysis, informing our ongoing development of Sigma Worksheet.

%% file: discussion.tex
\section{Discussion}\label{sec:discussion}
Sigma Worksheet has enabled our business users to successfully
use the data warehouse to perform ad-hoc analysis. Giving
such power to users has also raised new challenges and opportunities.


\subhead{Data Discovery} Finding the right tables in the database
remains a challenge for some users---this is important as Worksheet is only useful once a table is selected. In practice,
some companies have multiple databases connected to Sigma, each of
which can contain hundreds or thousands of tables. The addition of
data discovery functionality, implemented as search and
recommendations stands out as a valuable addition to the Sigma product.

\ifarxiv{\vspace{3ex}}\fi
\subhead{Spreadsheet Semantics} Our user feedback  indicates that
some users expect the Sigma Worksheet to behave more like an actual spreadsheet. These users are surprised when they cannot perform common
operations like editing and referencing individual cells or adding and
removing rows from the data table. There are also subtle differences
in our calculation semantics---for example spreadsheets and SQL
databases have different behaviors around \sql{NULL} or empty cells.
As such, there are many opportunities to further extend the Worksheet interface in order to meet the expectations of spreadsheet users.

\subhead{Power of Examples} Although Worksheet supports re-use through composition and parameterization, a more primitive mechanism, copying
and then editing another user's Worksheet, has turned out to be an important workflow in Sigma. It has made analysis accessible for users who otherwise wouldn't know how to get started. Making it easier to create and
share analysis examples and snippets for the Sigma Worksheet user community is an important future work.

\subhead{Expressivity} While Sigma Worksheet is able to express a number of complex analytical operations, it can be further extended to
support additional operations such as a function that is equivalent to Excel's \excel{VLOOKUP}~\cite{vlookup}.
Another topic to investigate is enabling aggregate and window calculations that do not rely on the Worksheet's levels. While removing the need to use an explicit \sql{GROUP BY} clause is a usability improvement, at times it might lead to analyses with convoluted  user interface configurations.

\subhead{Hybrid Evaluation}
Sigma Worksheet is capable of expressing complex analysis
that would be impractical or impossible to write by hand. This can lead to generated SQL queries that strain the optimization and execution capabilities of databases. While better query generation can reduce
of this stress, we cannot directly control how databases perform as data and/or user load increase or available compute resources decrease. Existing products have addressed this by building specialized query processing systems to guarantee interactivity - we believe this approach is still appropriate. Since Sigma Worksheet's UI is delivered through a web browser, we see opportunities for innovation through in-browser query processing.

\ignore {

\section{Lessons Learned}\label{sec:lessons}

The Worksheet has enabled our business users to successfully use the
data warehouse to perform ad-hoc analysis. However giving such power
has raised new issues:

User success is conditional on their understanding of the warehouse
schema. Further work is required to enable users to navigate
potentially thousands of tables in a typical warehouse.

The interactive experience of Sigma Worksheet rests on timely
responses to database queries. When the database is slow, our users
need a framework to reason about why so that they can remedy or avoid
the slowness.

Although the Worksheet supports re-use through composition and parameterization,
a more primitive mechanism, copying and then editing another user's Worksheet,
has turned out to be an import workflow in Sigma. It has made the database
accessible for users who otherwise wouldn't know how to get started.
}

%% file: related.tex
\section{Related Work} \label{sec:related}

The Worksheet draws from prior work on data cubes, business
intelligence (BI) tools, and spreadsheet systems.

\subhead{Data Cubes} The data cube is a relational operator that
enables the expression of multiple aggregate calculations in a single
query. One of the motivations behind its development was to address
the problems caused by applications generating complex and
difficult-to-optimize queries~\cite{gray1997data}. This resulted in
the extension of SQL's \sql{GROUP~BY} with the \sql{ROLLUP},
\sql{CUBE}, and \sql{GROUPING SETS} operations, simplifying the
calculation of subtotals and crosstabs along dimensions of the
aggregation. The data cube with its relatively simple semantics has
had a large impact on the technologies and products used for
analytical query processing. The data cube is now the core data model
for many commercial analytics applications, and has practically become
synonymous with online analytical processing
(OLAP)~\cite{chaudhuri1997overview}.



While it is more powerful than a SQL \code{GROUP BY}, additional
subqueries are needed to perform operations such as joins,
projections, selections, or window calculations. This is sufficient
for many interactive visualization systems, however it is limited in
expressing the complex calculation graphs frequently found in
spreadsheets. The Worksheet query model builds upon data cubes through
its ability to express complex computations within the configuration
of a single relational operator. We use the combination of levels,
calculations, and filters to construct a spreadsheet-like calculation
graph that can express joins, aggregations, projections, selections,
and window calculations, all with a fraction of the specification
complexity of the corresponding SQL.

\subhead{Business Intelligence Tools} Business intelligence (BI) tools
aim to help enterprise users such as managers, salespeople, marketers,
and business analysts make sense of data residing in company databases
for improved decision making. The design of Sigma Worksheet is greatly
influenced by modern BI tools such as Tableau~\cite{tableau}, Power
BI~\cite{powerbi}, and Microstrategy~\cite{microstrategy}. These
systems are designed to leverage the advantages of the data cube
model, enabling users to easily and efficiently perform aggregate
calculations by slicing and dicing the dimensions of the cube to
calculate and visualize aggregate data.

Existing BI tools can generate queries that contain the cube
aggregation, along with basic projection and selection operations. In
some advanced cases they can also apply post-aggregate window
calculations such as ranking, or perform percent of total calculations
using subtotals. However, if the user cannot express their analysis
with this fixed order of operations~\cite{tableau:ooo}, then they must
fallback to handwritten queries, or interactive data preparation in
another interface or product~\cite{tableau:prep,trifacta}.

In contrast, Worksheet does not have a fixed order of operations.
Instead the order of operations is generated dynamically during the
compilation and generation of a SQL query. As such, the complexity of
a given worksheet scales with the needs of the analyst, rather than
forcing a fallback to another mode of query authoring.

\subhead{Spreadsheet Systems} Spreadsheets are one of the most
ubiquitous tools used by individuals who need to collect, manipulate,
and analyze data. They present users with an addressable data grid and
a flexible query model, which make data entry, experimentation, and
iteration on small datasets easy. They also effectively complement relational databases, which trade some of the
flexibility and ease of use for scalability, performance, and
reliability. This begs the popular
question~\cite{bakke2011schema,bendre2018towards,rahman2020benchmarking}:
how can we get the best of both worlds?

General purpose spreadsheets like Microsoft Excel and Google Sheets
have some support for querying external databases, as a means to
bridge the gap between the two worlds. Recent tools such as
Airtable~\cite{airtable}, Coda~\cite{coda}, and Notion~\cite{notion}
integrate some basic spreadsheet and database functionalities.
However, these systems lack the scale, reliability, and expressivity
of modern databases. At the time of writing, Excel is limited to
roughly 1M rows and 16K columns---by comparison the maximum length for
a single string in the Snowflake data warehouse is 16MB.

Unlike most other spreadsheet or spreadsheet-like tools, Worksheet
does not persist the content of the data grid. Since we rely on query
generation and execution in an external data warehouse, our query
model is able to leverage result pagination to visualize the results
of queries that produce thousands or millions of records. Furthermore
Worksheet presents users with a spreadsheet-like visual interface and
straightforward formulas, and relies on our query compiler to use the
levels, calculations, and filters to drive the generation of complex,
expressive SQL queries.

\cagatay{Revise to include a more up-to-date discussion of related
  work and ideas.} To the best of our knowledge, the only other system
that effectively combines the spreadsheet model with interactive
visualization of large databases is Sieuferd~\cite{Bakke:2016}. The
Sieuferd system is a standalone tool for direct query construction,
with a focus on relational completeness. In contrast, Worksheet is an
important component of a larger Software-as-a-Service (SaaS) analytics
product, and is focused on the construction of efficient and
optimizable OLAP queries.

%% file: conclusion.tex
\section{Conclusion}\label{sec:conclusion}
We have described Sigma Worksheet, a new interactive system designed
for effective visual analysis of enterprise data stored in cloud data
warehouses. Sigma Worksheet enables business users as well as analysts
to work with the full scale of their data and perform iterative ad-hoc
analysis without needing to write SQL queries. Simultaneously, Sigma
Worksheet empowers these users with versatility and expressivity built
on SQL and with accessibility enabled by an easy-to-use interface akin
to spreadsheet systems.
We see numerous opportunities ahead to extend the guided analysis,
data transformation, and SQL capabilities of Worksheet. Reconciling
these with performance and usability remains a challenge.


%% file: ack.tex
\section{Acknowledgments}
We thank everyone on the engineering and design team at Sigma
Computing. Sigma Worksheet is a product of their labor, care, and
dedication. We also thank the early users of Sigma for teaching us so
much and having patience with our bugs. Special thanks to Zack~Norton
for producing and narrating our demonstration video.

\ignore{
\section{Acknowledgments}
The authors wish to thank everyone on the engineering and design team
at Sigma Computing for their labor, care, and dedication to building
an exceptional product. We would also like to thank the early users of
Sigma for teaching us so much and having patience with our bugs.
Finally we would like to thank Sigma's Founders, Rob Woollen and Jason Frantz for their support and advice in writing this paper.
}